\documentclass[aps, superscriptaddress, longbibliography, notitlepage, twocolumn, floatfix]{revtex4-1}
\usepackage{amsmath,amssymb}
\usepackage{graphicx}
\usepackage{hyperref}
\usepackage{enumitem}
\usepackage{mathtools}
\usepackage{hypcap}
\usepackage{color}

\begin{document}

\title{Dynamics of cold random hyperbolic graphs with link persistence}

\author{Sofoclis Zambirinis}
\affiliation{Department of Electrical Engineering, Computer Engineering and Informatics, Cyprus University of Technology, 3036 Limassol, Cyprus}
\author{Harrison Hartle}
\affiliation{Network Science Institute, Northeastern University, Boston, Massachusetts 02115, USA}
\author{Fragkiskos Papadopoulos}
\email{f.papadopoulos@cut.ac.cy}
\affiliation{Department of Electrical Engineering, Computer Engineering and Informatics, Cyprus University of Technology, 3036 Limassol, Cyprus}

\date{\today}

\begin{abstract}
We consider and analyze a dynamic model of random hyperbolic graphs with link persistence. In the model, both connections and disconnections can be propagated from the current to the next snapshot with probability $\omega \in [0, 1)$. Otherwise, with probability $1-\omega$, connections are reestablished according to the random hyperbolic graphs model. We show that while the persistence probability  $\omega$ affects the averages of the contact and intercontact distributions, it does not affect the tails of these distributions, which decay as power laws with exponents that do not depend on $\omega$. We also consider examples of real temporal networks, and we show that the considered model can adequately reproduce several of their dynamical properties. Our results advance our understanding of the realistic modeling of temporal networks and of the effects of link persistence on temporal network properties.
\end{abstract}

\maketitle

\section{Introduction}

Random hyperbolic graphs (RHGs) have been shown to be adequate for modeling real complex networks, as they naturally and simultaneously possess many of their common structural characteristics. Such characteristics include heterogeneous degree distributions, strong clustering, and the small-world property~\cite{Krioukov2009, Krioukov2010, Panagiotou2012, boguna2020, fountoulakis2021}. RHGs are adequate only in the ``cold regime," where the \emph{network temperature} $T$ in the model takes values between $0$ and $1$. This is because only when $T \in [0, 1)$ can RHGs have strong clustering, as observed in real systems~\cite{Krioukov2010}.  Cold RHGs have been successfully used as a basis in maximum likelihood estimation methods that infer the hyperbolic node coordinates in real systems, facilitating important applications that include community detection, missing and future link prediction, network navigation, and network dismantling~\cite{Boguna2010, frag:hypermap, frag:hypermap_cn, kleineberg2016, GarciaPerez2019, Serrano2011, ndn2016, allard2020, osat2022}.

Recently, the simplest possible version of dynamic RHGs, the \emph{dynamic-$\mathbb{S}^1$} model, has been proposed and analyzed~\cite{Papadopoulos2019}. In the dynamic-$\mathbb{S}^1$, the hyperbolic node coordinates remain fixed, while each network snapshot $G_t$ is constructed anew using the static $\mathbb{S}^1$ model, or equivalently, the hyperbolic $\mathbb{H}^2$ model~\cite{Krioukov2010}. It has been shown that the dynamic-$\mathbb{S}^1$ can qualitatively (and some times quantitatively) reproduce many \emph{temporal} network properties observed in real systems, such as the broad distributions of contact and intercontact durations and the abundance of recurrent components~\cite{Papadopoulos2019, flores2018}. 

Correlations among the network snapshots in the dynamic-$\mathbb{S}^1$ are imposed by the nodes' hyperbolic coordinates; nodes at smaller hyperbolic distances
have higher chances of being connected in each snapshot, intuitively explaining why heterogeneous (inter)contact distributions emerge in the model.  In particular, the contact and interconnect distributions are power laws in the model, with respective exponents $2+T \in (2, 3)$ and $2-T \in(1, 2)$~\cite{Papadopoulos2019}. These distributions are remarkably consistent with (inter)contact distributions observed in some real systems. For instance, in human proximity networks, studies have reported power-law contact distributions with exponents larger than or close to $2$~\cite{Scherrer2008, SPcontactexp}, and power-law intercontact distributions with exponents between $1$ and $2$~\cite{hui_paper, chaintreau_paper, Partners2011, HighSchoolData2}. Based on the dynamic-$\mathbb{S}^1$, human proximity networks have been recently mapped to hyperbolic spaces, and related applications have been explored~\cite{flores2020}. We note that the dynamic-$\mathbb{S}^1$ exhibits realistic dynamical properties only in the cold regime ($T \in(0, 1)$) but not in the hot ($T > 1$)~\cite{Papadopoulos2022}.

In this paper, we observe that synthetic temporal networks constructed with the dynamic-$\mathbb{S}^1$ may underestimate the average contact and intercontact durations in the corresponding real systems. This observation suggests that in addition to purely geometric aspects the explicit link formation process in one snapshot may impact the topology of subsequent snapshots in real networks. Motivated by this observation, we consider and analyze a generalization of the dynamic-$\mathbb{S}^1$ with \emph{link persistence}~\cite{mazzarisi2020, Papadopoulos2019lp, hartle2021}, called $\omega$-dynamic-$\mathbb{S}^1$. In the $\omega$-dynamic-$\mathbb{S}^1$, both connections and disconnections can persist, i.e., propagate, from the current to the next snapshot with probability $\omega \in [0, 1)$. Otherwise, with probability $1-\omega$, connections are reestablished according to the $\mathbb{S}^1$ model. The case $\omega=0$ corresponds to the dynamic-$\mathbb{S}^1$~\cite{Papadopoulos2019}. 

 We perform a detailed mathematical analysis of the contact and intercontact distributions in the $\omega$-dynamic-$\mathbb{S}^1$. One of our main results is that while the persistence probability $\omega$ affects the averages of the (inter)contact distributions, it does not affect the tails of these distributions. Specifically, we show that for sufficiently sparse networks the (inter)contact distributions decay as power laws with the same exponents as in the dynamic-$\mathbb{S}^1$. We also show that synthetic networks constructed with the $\omega$-dynamic-$\mathbb{S}^1$ can reproduce several dynamical properties of real systems, while better capturing their average (inter)contact durations. These results advance our understanding of realistically modeling of temporal networks and of the effects of link persistence. In particular, our results suggest that link persistence in real systems may affect only the averages but not the tails of the (inter)contact distributions, which are important properties affecting the capacity and delay of a network and the dynamics of spreading processes~\cite{ContiOppurtunisticOverview, vazquez2007, timo2009, Karsai2011, machens2013, gauvin2013}. For instance, it has been shown that heterogeneous inter-event distributions may slow down epidemic spreading~\cite{vazquez2007, Karsai2011}. Since link persistence does not affect the tail of the intercontact distribution, it may not affect the characteristics of related epidemic spreading measures~\cite{vazquez2007, Karsai2011}.
 
 Intuitively, a higher persistence for non-links means that nodes will tend to stay disconnected for a longer period of time, which can slow down epidemic spreading. This slow-down could be more important for intercontacts that would otherwise be short, e.g., intercontacts between more similar nodes. On the other hand, a higher persistence for links means that nodes will tend to stay connected for a longer period of time, which can increase the chances of transmitting a communicable disease. This effect could be more important for contacts that would otherwise be short, e.g., contacts between less similar nodes. Investigating the exact effects of link persistence on epidemic spreading is an interesting avenue for future work.
 
 The rest of the paper is organized as follows. In the next section we provide an overview of the $\mathbb{S}^{1}$ model. In Sec.~\ref{sec:omega_dynamic_S1} we present the $\omega$-dynamic-$\mathbb{S}^1$. In Sec.~\ref{sec:real_vs_modeled} we illustrate that the $\omega$-dynamic-$\mathbb{S}^1$ can reproduce several dynamical properties of real networks, while acurrately capturing their average contact durations. In Sec.~\ref{sec:analysis} we perform a detailed mathematical analysis of the contact and intercontact distributions in the $\omega$-dynamic-$\mathbb{S}^1$. Furthermore, we analyze the expected time-aggregated degree in the model. In Sec.~\ref{sec:related_work} we discuss other relevant work. Finally, we conclude the paper with a discussion and future work directions in Sec.~\ref{sec:conclusion}. 

\section{$\mathbb{S}^{1}$ model}
\label{sec:S1}

In the $\mathbb{S}^{1}$ model~\cite{Krioukov2010} each node has latent (or hidden) variables $\kappa$ and $\theta$. The latent variable $\kappa$ is proportional to the node's expected degree in the resulting network and abstracts its \emph{popularity}. The latent variable $\theta$ is the angular \emph{similarity} coordinate of the node on a circle of radius $R=N/2\pi$, where $N$ is the total number of nodes~\cite{Papadopoulos2012}. To construct a network with the model that has size $N$, average node degree $\bar{k}$, and temperature $T \in (0,1)$, we perform the following steps:
\begin{enumerate}
\item[(1)] coordinate assignment: for each node $i=1, 2,\ldots,N$, sample its angular coordinate $\theta_i$ uniformly at random from $[0, 2\pi]$, and its degree variable $\kappa_i$ from a probability density function (PDF) $\rho(\kappa)$;
\item[(2)] creation of edges: connect every pair of nodes $i, j$ with the Fermi-Dirac connection probability
\begin{align}
\label{eq:p_s1}
p_{ij}=\frac{1}{1+\chi_{ij}^{1/T}}.
\end{align}
\end{enumerate}
In the last expression, $\chi_{ij}$ is the effective distance between nodes $i$ and $j$,
\begin{align}
\label{eq:chi}
\chi_{ij} = \frac{R \Delta\theta_{ij}}{\mu \kappa_i \kappa_j},
\end{align}
where $\Delta \theta_{ij}=\pi - | \pi -|\theta_i - \theta_j||$ is the similarity distance between $i$ and $j$. We note that since $\theta$ is uniformly distributed on $[0, 2\pi]$, the PDF of $\Delta \theta$ is the uniform PDF on $[0, \pi]$, $f(\Delta \theta)=1/\pi$. 

Parameter $\mu$ in~(\ref{eq:chi}) is derived from the condition that the expected degree in the network is indeed $\bar{k}$. For sparse networks ($N \gg \bar{k}$)
\begin{align}
\label{eq:mu}
\mu=\frac{\bar{k}\sin{(T \pi)}}{2\bar{\kappa}^2 T \pi},
\end{align}
where $\bar{\kappa} = \int \kappa \rho(\kappa) \mathrm{d} \kappa$.  Further, the expected degree of a node with latent variable $\kappa$ can be computed as
\begin{align}
\label{eq:kappa}
\bar{k}(\kappa)= \frac{\bar{k}}{\bar{\kappa}}\kappa \propto \kappa.
\end{align}
For sparse networks, the resulting degree distribution $P(k)$ has a similar functional form as $\rho(\kappa)$~\cite{Boguna2003}.  We also note that smaller values of  the temperature $T$ favor connections at smaller effective distances and increase the average clustering~\cite{Dorogovtsev10-book} in the network, which is maximized at $T \to 0$.  

The $\mathbb{S}^{1}$ model is equivalent to RHGs, i.e., to the hyperbolic $\mathbb{H}^{2}$ model~\cite{Krioukov2010}, after a simple transformation of the degree variables $\kappa$ to radial coordinates $r$ on the hyperbolic disk. See Ref.~\cite{Krioukov2010} for further details.  

\section{$\omega$-dynamic-$\mathbb{S}^{1}$}
\label{sec:omega_dynamic_S1}

The $\omega$-dynamic-$\mathbb{S}^{1}$ models a sequence of network snapshots, $G_t$, $t=1,\ldots, \tau$, where $\tau$ is the total number of time slots. In the model there are $N$ nodes that are assigned latent variables $\kappa, \theta$ as in the $\mathbb{S}^{1}$ model, which remain fixed in all time slots. The temperature $T$ and the persistence probability $\omega$ are also fixed, while each snapshot $G_t$ is allowed to have a different average degree $\bar{k}_t$. Thus, the model parameters are $N, \tau, \rho(\kappa),T, \omega$, and $\bar{k}_t, t=1, \ldots, \tau$. 

Let
\[
e_{ij}^{(t)} = \begin{cases}
	1  &  \textrm{if nodes } (i,j) \textrm{ are connected at time } t, \\
	0  &  \textrm{otherwise}. 
\end{cases}
\]
The snapshots in the $\omega$-dynamic-$\mathbb{S}^{1}$ are generated according to the following simple rules:
\begin{enumerate}
\item[(1)] snapshot $G_1$ is a realization of the $\mathbb{S}^{1}$ model with average degree $\bar{k}_1$;
\item[(2)] at each time step $t=2, \ldots, \tau$, snapshot $G_t$ starts with $N$ disconnected nodes and has target average degree $\bar{k}_t$;
\item [(3)] each pair of nodes $i, j$ in snapshot $G_t$ connects according to the following conditional connection probabilities:
\begin{align}
\label{eq:cp1}
P[ e_{ij}^{(t)} = 1  |  e_{ij}^{(t-1)} = 1 ]  &=  \omega+(1-\omega) p_{ij}^{(t)},\\
\label{eq:cp2}
P[ e_{ij}^{(t)} = 1  |  e_{ij}^{(t-1)} = 0 ]  &=  (1-\omega) p_{ij}^{(t)},
\end{align}
where $p_{ij}^{(t)}$ is given by~(\ref{eq:p_s1}), with $\bar{k}$ in~(\ref{eq:mu}) set equal to $\bar{k}_t$;
\item[(4)] at time $t+1$, the process is repeated to generate snapshot $G_{t+1}$.
\end{enumerate}
Equation~(\ref{eq:cp1}) is the case in which the node pair $i, j$ is connected in the previous time slot $t-1$. In that case, the pair is connected in slot $t$ either because the connection has been propagated from $t-1$ (with probability~$\omega$) or because the connection has been established according to $p_{ij}^{(t)}$ (with probability~$1-\omega$).  Equation~(\ref{eq:cp2}) is the case in which the pair $i, j$ is not connected in $t-1$. In that case, the pair can be connected in slot $t$ if the disconnection has not been propagated from $t-1$ (with probability $1-\omega$) and the pair connected according to $p_{ij}^{(t)}$. 

We note that the unconditional connection probability for node pair $i, j$ at time $t=2, 3, \ldots$, can be written as
\begin{align}
\label{eq:upcr}
\nonumber P[ e_{ij}^{(t)} = 1 ] &= P[ e_{ij}^{(t)} = 1  |  e_{ij}^{(t-1)} = 1 ] P[ e_{ij}^{(t-1)} = 1 ]\\
\nonumber & + P[ e_{ij}^{(t)} = 1  |  e_{ij}^{(t-1)} = 0 ] \{1-P[ e_{ij}^{(t-1)} = 1 ]\}\\
&= \omega P[ e_{ij}^{(t-1)} = 1 ] + (1-\omega) p_{ij}^{(t)}.
\end{align}
Solving the above recurrence equation yields
\begin{equation}
\label{eq:upc}
P[ e_{ij}^{(t)} = 1 ] = \omega^{t-1} p_{ij}^{(1)}+(1-\omega) \sum_{s=0}^{t-2} \omega^s p_{ij}^{(t-s)}.
\end{equation}
Notice that if each snapshot has the same average degree, $\bar{k}_t = \bar{k}, \forall t$, then $p_{ij}^{(t)}$ is the same in all slots, $p_{ij}^{(t)}=p_{ij}, \forall t$, and~(\ref{eq:upc}) simplifies to
\begin{equation}
P[ e_{ij}^{(t)} = 1 ] = p_{ij}.
\end{equation}
In other words, if $\bar{k}_t = \bar{k}, \forall t$, then the unconditional connection probability is exactly the same as the connection probability in the $\mathbb{S}^{1}$ model. Thus, as a side note, in this case the $\omega$-dynamic-$\mathbb{S}^{1}$ satisfies the \emph{equilibrium property}, in the sense that individual snapshots in the model are indistinguishable from static-model realizations~\cite{hartle2021}. The equilibrium property is also satisfied for $\omega=0$, i.e., when there is no link persistence, in which case $P[ e_{ij}^{(t)} = 1 ] = p_{ij}^{(t)}$. Otherwise, the equilibrium property is not satisfied.

\section {Real~vs.~modeled networks}
\label{sec:real_vs_modeled}

\subsection{Real networks}
\label{sec:real_nets}

To illustrate the realism of the model we compare its properties against the properties of five real temporal networks. These networks according to the model have a different  link-persistence probability~$\omega$. Specifically, we consider three face-to-face interaction networks from SocioPatterns~\cite{SocioPatterns}, which correspond to a high school in Marseilles~\cite{HighSchoolData}, a primary school in Lyon~\cite{PrimaryData}, and a village in rural Malawi~\cite{MalawiData}.  These networks were captured over a period of $5$, $2$ and $13$ days, respectively. Each of their snapshots corresponds to a slot of $20$ s. 

Further, we consider the network of coded interactions between socio-political actors from the Integrated Crisis Early Warning System (ICEWS)~\cite{ICEWS}, as well as the e-mail communication network between members of a European research institution (Email-EU)~\cite{email, emailpaper}.  We consider 401 daily snapshots of the ICEWS network (days $3000$-$3400$ in the data). For the Email-EU network we consider only bidirectional communications corresponding to $79$ weekly snapshots (from October 2003 to May 2005). In all cases we number the time slots and assign node IDs sequentially, $t=1,2,\ldots, \tau$ and $i=1,2, \ldots, N$. Table~\ref{tableReal} gives an overview of the data.
 
\subsection{Modeled counterparts}
\label{sec:modeled_nets}

For each real network we construct its synthetic counterpart using the $\omega$-dynamic-$\mathbb{S}^{1}$, following a similar procedure to that in Ref.~\cite{Papadopoulos2019}. Specifically, each counterpart has the same number of nodes $N$ and duration $\tau$ as the corresponding real network, while the latent variable $\kappa_i$ of each node $i=1, \ldots, N$ is assigned as follows. First, for each real network we compute the average degree per slot of each node $i$,
\begin{equation}
\label{eq:d_i_obs}
\bar{d}_i=\frac{1}{\tau}\sum_{t=1}^{\tau} d_{i, t},
\end{equation}
where $d_{i,t} \geq 0$ is node's $i$ degree in slot $t$. Then, we set
\begin{equation}
\label{eq:eq1}
\kappa_i=\bar{d}_i.
\end{equation} 
The angular coordinate $\theta_i$ of each node $i$ is sampled uniformly at random from $[0, 2 \pi]$. Further, the target average degree $\bar{k}_t$ in each snapshot $G_t$, $t=1, \ldots, \tau$, is set equal to the average degree in the corresponding real snapshot at slot $t$,
\begin{equation}
\label{eq:k_t}
\bar{k}_t=\frac{1}{N}\sum_{i=1}^{N} d_{i, t}.
\end{equation}
Finally, the temperature $T$ and the link-persistence probability $\omega$ are simultaneously tuned such that the resulting average time-aggregated degree $\bar{k}_\textnormal{aggr}$ and the average contact duration $\bar{t}_\textnormal{c}$ are similar to the ones in the real network. We perform this tuning manually, by running simulations with different values of $T$ and $\omega$ until we find the combination that produces similar values for $\bar{k}_\textnormal{aggr}$ and $\bar{t}_\textnormal{c}$ as in the corresponding real network. Fig.~\ref{fig:ave_contact} in Sec.~\ref{sec:analysis_contact} illustrates the dependence of $\bar{t}_\textnormal{c}$ on both $T$ and $\omega$, while Fig.~\ref{fig:kaggr_val} in Sec.~\ref{sec:kaggr} shows how $\bar{k}_\textnormal{aggr}$ depends on these parameters. The values of $T$ and $\omega$ that we find for each case are reported in Table~\ref{tableSimulated}. (We note that we do not explicitly match the average intercontact duration $\bar{t}_\textnormal{ic}$ in each real network.)

\subsection{Properties of modeled~vs.~real networks}
\label{sec:properties}

\begin{table}
\begin{tabular}{|c|c|c|c|c|c|c|c|c|c|}
\hline 
Real network & $N$ & $\tau$ & $\bar{n}$  &  $\bar{k}$ & $\bar{k}_\textnormal{aggr}$ & $\bar{t}_\textnormal{c}$ & $\bar{t}_\textnormal{ic}$\\ 
\hline 
High School & 327 & 18179 & 17 &  0.06 & 36& 2.79 & 527 \\ 
\hline 
Primary School & 242 & 5846 & 30 &  0.18 & 69 &1.62 & 229\\ 
\hline 
Malawi Village & 86 & 57791 & 3.4 &  0.04 & 8.1 & 2.91 & 213\\ 
\hline 
ICEWS & 29047 & 401 & 1089 &  0.09 & 13 & 1.19 & 40\\ 
\hline
Email-EU & 980 & 79 & 549 &  2.96 & 33 & 1.84 & 6.5\\ 
\hline
\end{tabular}
\caption{Overview of the considered real networks. $N$ is the total number of nodes seen; $\tau$ is the total number of time slots; $\bar{n}$ is the average number of interacting (i.e., nonzero degree) nodes per slot; $\bar{k}$ is the average snapshot degree ($\bar{k}=\sum_{t=1}^{\tau} \bar{k}_t/\tau$); $\bar{k}_\textnormal{aggr}$ is the average degree in the time-aggregated network, i.e., the average number of other nodes that a node connects to at least once;  $\bar{t}_\textnormal{c}$ is the average contact duration, i.e., the average number of consecutive slots in which two nodes remain connected; and  $\bar{t}_\textnormal{ic}$ is the average intercontact duration, i.e., the average number of consecutive slots in which two nodes remain disconnected. Average values above $10$ are rounded to the nearest integer.
\label{tableReal}}
\bigskip
\begin{tabular}{|c|c|c|c|c|c|c|c|c|c|}
\hline 
Modeled network & $\bar{n}$ & $\bar{k}$ & $\bar{k}_\textnormal{aggr}$ & $\bar{t}_\textnormal{c}$ & $\bar{t}_\textnormal{ic}$ & $T$ & $\omega$ \\  
\hline 
High School & 18 & 0.06 & 36 & 2.75 & 482 & 0.67 & 0.46\\ 
\hline 
Primary School & 32 & 0.16 & 69 &1.66 & 246 & 0.75 & 0.18\\ 
\hline 
Malawi Village & 3.4 & 0.05 & 7.9 & 2.90 & 228 & 0.45 & 0.38\\ 
\hline 
ICEWS & 1033 & 0.08 & 12 & 1.17  & 36 & 0.90 & 0\\ 
\hline 
Email-EU & 592 & 3.34 & 36 & 1.86 & 6.8 & 0.53 & 0\\ 
\hline 
\end{tabular}
\caption{Modeled counterparts. Average values correspond to averages across $100$ simulation runs, except for the ICEWS network, where the values are results from one run. Each counterpart has the same number of nodes $N$ and duration $\tau$ as the corresponding real network in Table~\ref{tableReal}.  
\label{tableSimulated}}
\end{table}

\begin{figure*}
\includegraphics[width=18cm, height=3.2cm]{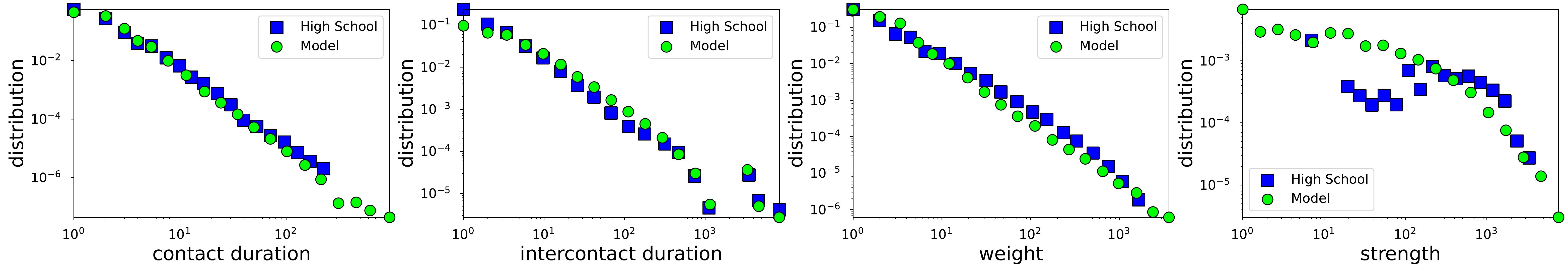}
\includegraphics[width=18cm, height=3.2cm]{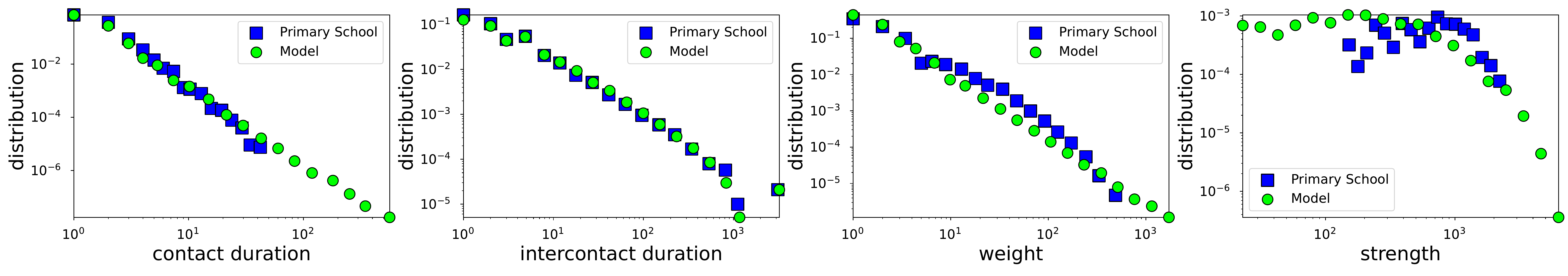}
\includegraphics[width=18cm, height=3.2cm]{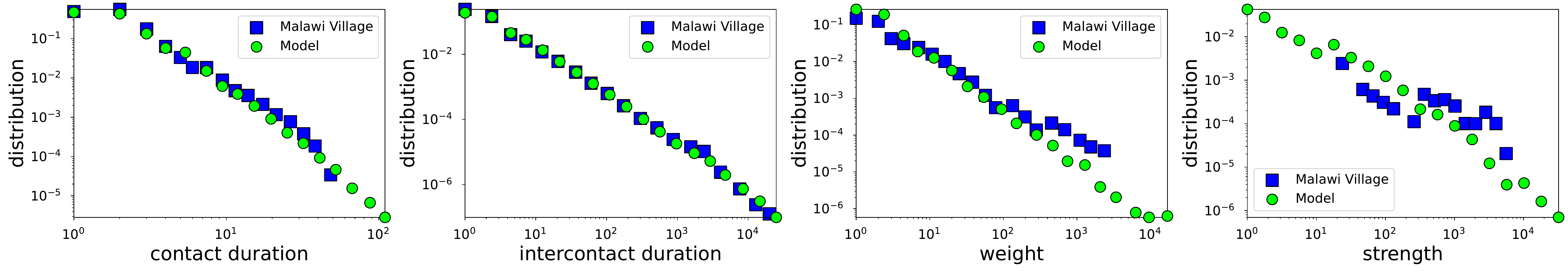}
\includegraphics[width=18cm, height=3.2cm]{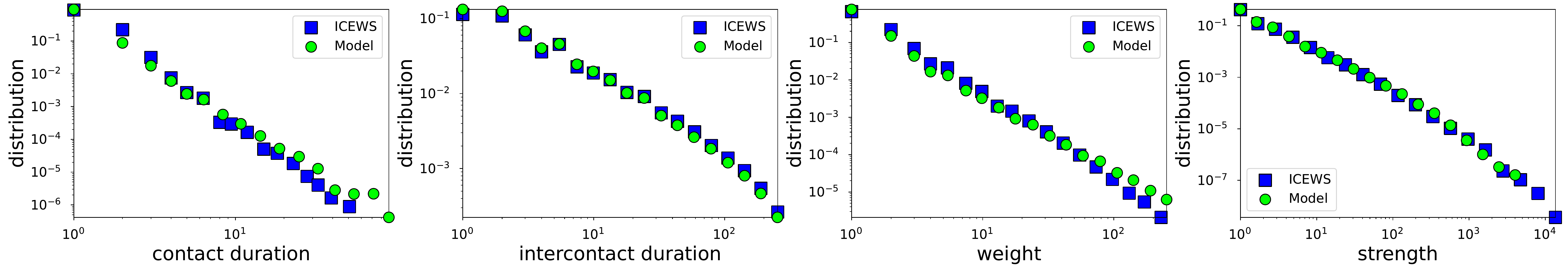}
\includegraphics[width=18cm, height=3.2cm]{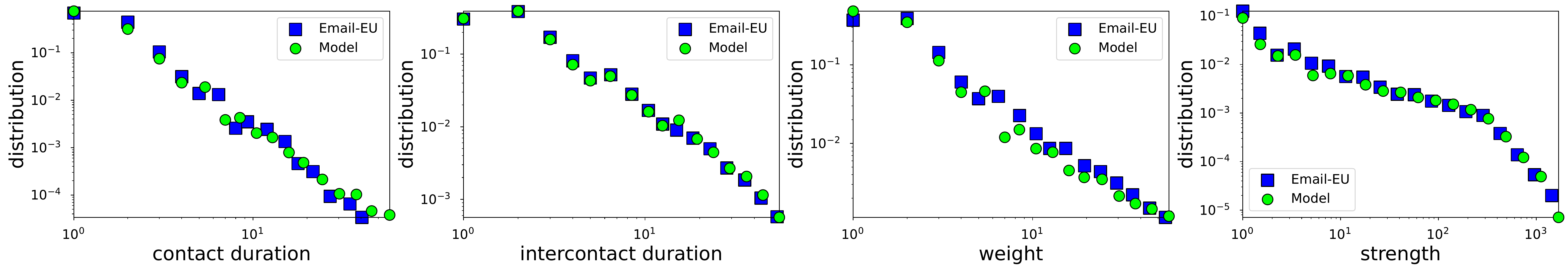}
\caption{Real~vs.~modeled networks. The plots show the contact distributions (first column), intercontact distributions (second column), weight distributions (third column), and strength distributions (fourth column). The values in the $y$-axes of the plots represent relative frequencies, i.e., they are computed as $n_t/\sum_{j} n_j$, where $n_t$ is the number of samples that have value $t$. All plots have been binned logarithmically. The results with the model are averages over five simulation runs, except for the ICEWS network, where the results are from one run. Durations are measured in numbers of days and weeks for the ICEWS and Email-EU networks. For the other networks they are measured in numbers of slots of 20 s.
\label{figAll}}
\end{figure*}

\begin{figure*}
\includegraphics[width=3.5cm, height=2.8cm]{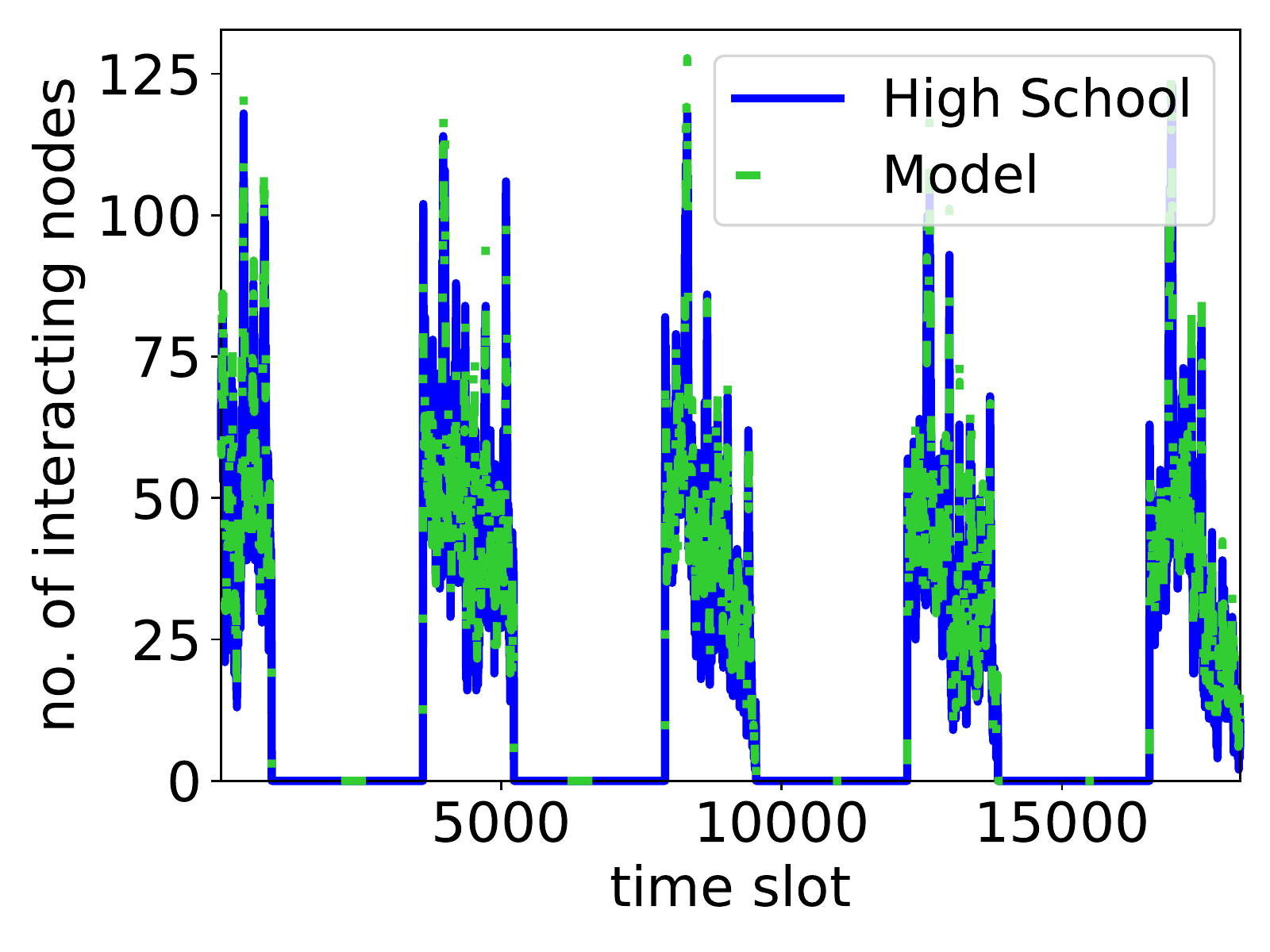}
\includegraphics[width=3.5cm, height=2.8cm]{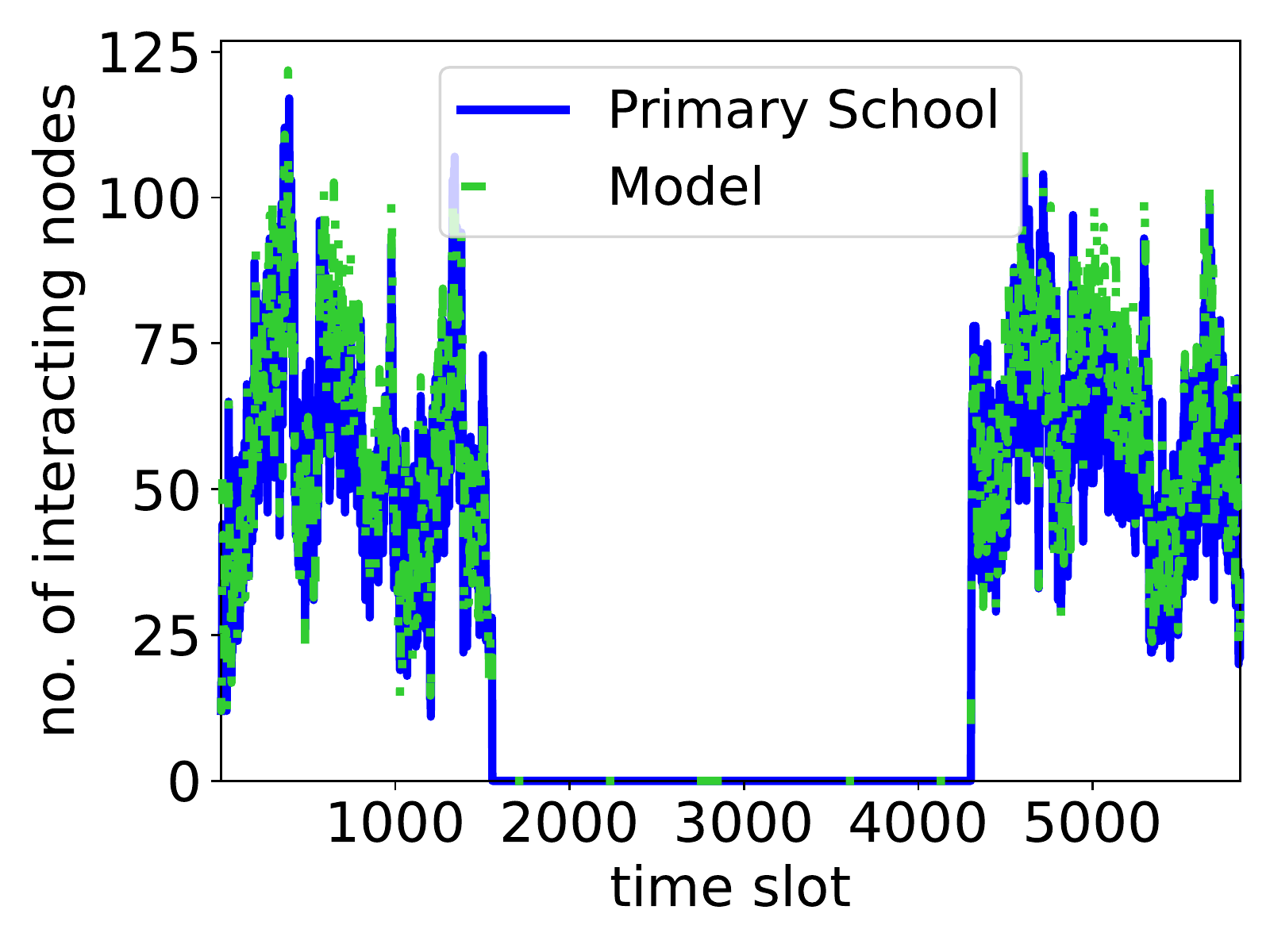}
\includegraphics[width=3.5cm, height=2.8cm]{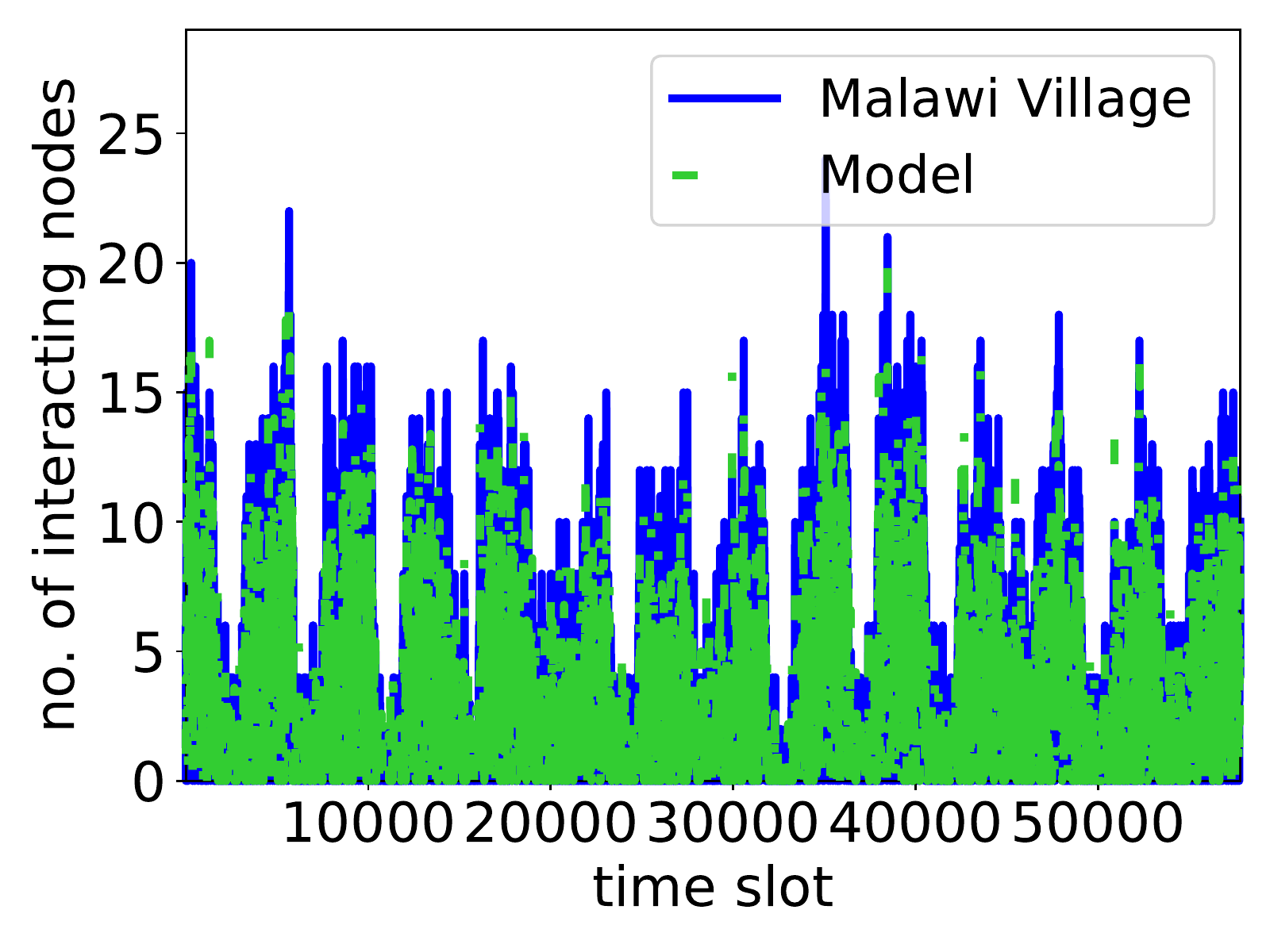}
\includegraphics[width=3.5cm, height=2.8cm]{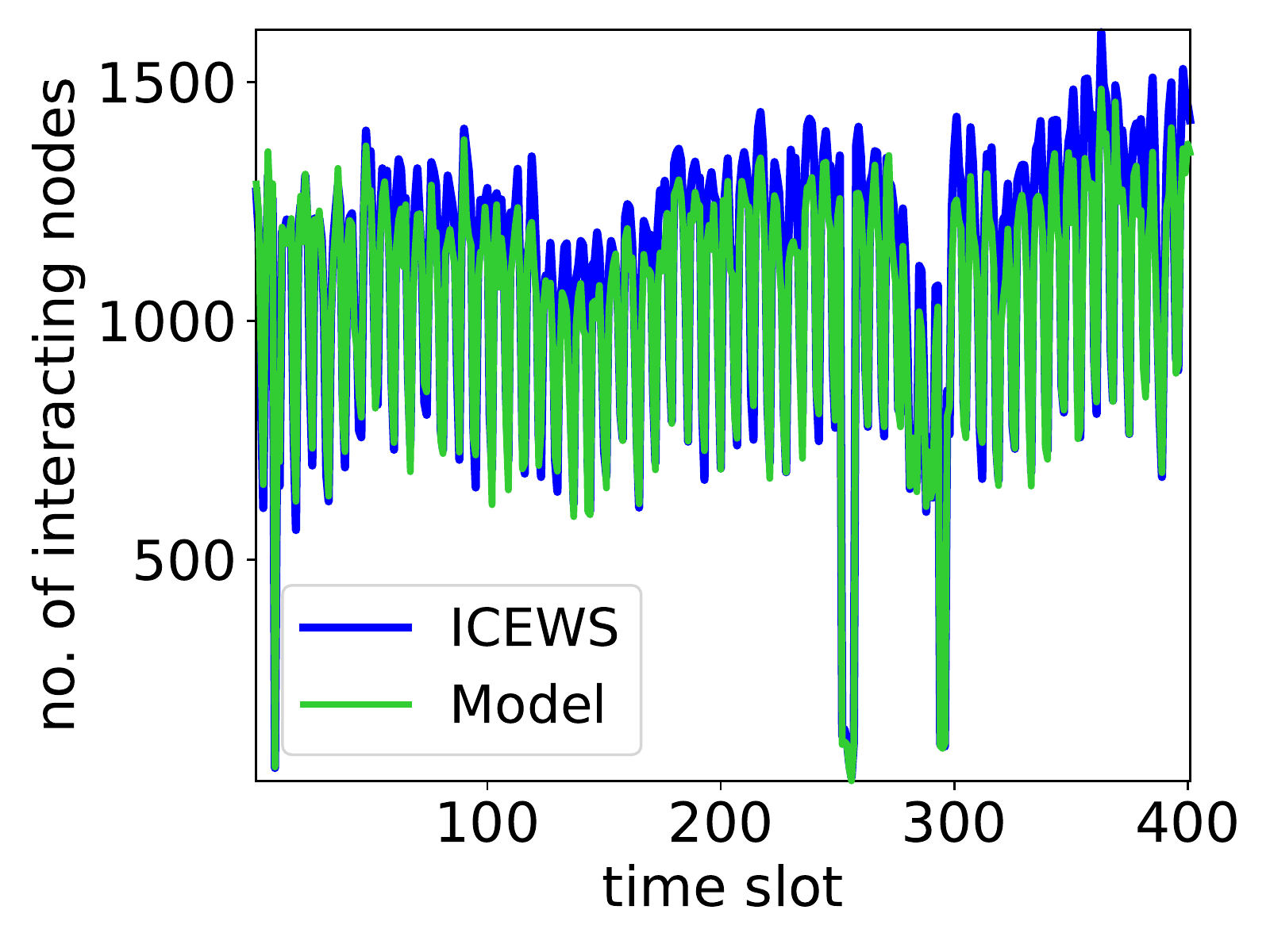}
\includegraphics[width=3.5cm, height=2.8cm]{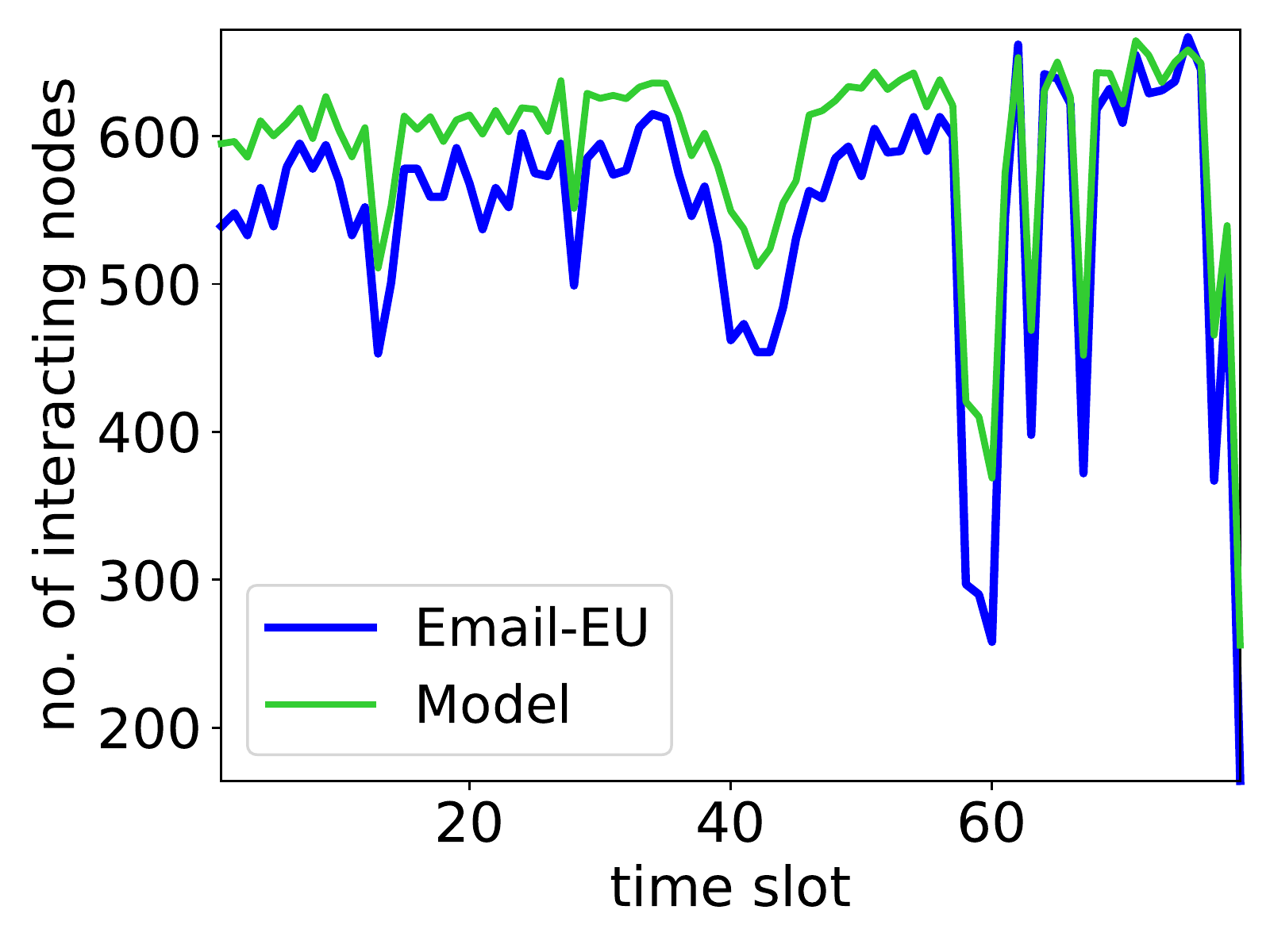}
\caption{Number of interacting nodes per time slot in real and modeled networks.
\label{fignodes}}
\end{figure*}

Table~\ref{tableSimulated} gives an overview of the modeled counterparts. We see that their characteristics are overall very similar to the ones of the real networks in Table~\ref{tableReal}. 
Further, in Fig.~\ref{figAll} we also compare the following properties between real and modeled networks:
\begin{itemize}[noitemsep]
\item[(a)] The contact distribution, which is the distribution of the number of consecutive slots in which a pair of nodes remains connected.
\item[(b)] The intercontact distribution, which is the distribution of the number of consecutive slots in which a pair of nodes remains disconnected.
\item [(c)] The weight distribution, which is the distribution of the edge weights in the time-aggregated network. In the time-aggregated network, two nodes are connected if they were connected in at least one slot, while the edge-weight in this network is the total number of slots in which the two end points of the edge were connected. 
\item[(d)] The strength distribution, which is the distribution of the node strengths in the time-aggregated network. The strength of a node is the sum of the weights of all edges attached to the node. 
\end{itemize}
Fig.~\ref{figAll} shows that the modeled counterparts capture all the above properties in the real systems remarkably well. Further, Fig.~\ref{fignodes} shows that the counterparts can also capture the variability of the number of interacting nodes per slot. The model can also capture several other properties of the considered real systems, as in Ref.~\cite{Papadopoulos2019}, which we omit here for brevity.

We note that $\omega=0$ in the ICEWS and Email-EU counterparts, suggesting that there is no link persistence in the corresponding real systems. On the other hand, $\omega > 0$ in the counterparts of the considered face-to-face interaction networks. We note that one can model these systems using $\omega=0$ and still qualitatively reproduce their properties, cf.~\cite{Papadopoulos2019}, but the average contact and intercontact durations will be underestimated in that case. Specifically, the values of $\{\bar{t}_{\textnormal{c}}, \bar{t}_{\textnormal{ic}}\}$ in synthetic counterparts of the high school, primary school, and Malawi village networks, constructed as described in Sec.~\ref{sec:modeled_nets} but with $\omega=0$, are, respectively, $\{\bar{t}_{\textnormal{c}}=1.62, \bar{t}_{\textnormal{ic}}=298\}$, $\{\bar{t}_{\textnormal{c}}=1.41, \bar{t}_{\textnormal{ic}}=207\}$, and  $\{\bar{t}_{\textnormal{c}}=2.03, \bar{t}_{\textnormal{ic}}=167\}$ (versus the values in Tables~\ref{tableReal} and~\ref{tableSimulated}).

In the next section we focus on the contact and intercontact distributions in the $\omega$-dynamic-$\mathbb{S}^{1}$, and we prove their properties. We also analyze the expected time-aggregated degree in the model, elucidating its dependence on both the temperature $T$ and the link-persistence probability $\omega$.

\section{Analysis}
\label{sec:analysis}
 
To facilitate the analysis, we assume $\bar{k}_t = \bar{k}$,~$\forall t$, i.e., that all snapshots have the same average degree $\bar{k}$. This assumption renders the connection probability $p_{ij}$ in~(\ref{eq:p_s1}) the same in all time slots. However, we note that our analytical results follow closely the simulation results from the modeled counterparts of the previous section, where this assumption does not hold.

One of our main results is that for sufficiently sparse snapshots, $N \gg \bar{k}$, the contact and intercontact distributions decay as power laws with exponents $2+T$ and $2-T$, irrespective of the value of the persistence probability $\omega$. Technically, we consider these distributions in the limit $N \to \infty$. However, the same results also hold in the limit $N/\bar{k} \to \infty$, which includes the case when $N$ is finite and $\bar{k} \to 0$ (this case may be more relevant to some real networks, such as face-to-face interaction networks, where their size is relatively small but they are still sparse (cf.~Table~\ref{tableReal})). As it will become apparent, these results do not depend on the distribution of the expected node degrees, i.e., on $\rho(\kappa)$. We begin with the contact distribution. 

\subsection{Contact distribution}
\label{sec:analysis_contact}

Consider the probability to observe a sequence of exactly $t$ consecutive slots, where two nodes $i$ and $j$ with latent degrees $\kappa_i$ and $\kappa_j$ and angular distance $\Delta \theta_{ij}$ are connected, $t=1, 2, \ldots, \tau-2$. This probability, denoted by $r_\textnormal{c}(t ; \kappa_i, \kappa_j, \Delta\theta_{ij})$, is the percentage of observation time $\tau$ where we observe a slot where these two nodes are not connected, followed by $t$ slots where they are connected, followed by a slot where they are again not connected.

For each duration $t$, there are $\tau-t-1$ possibilities where this duration can be realized. For instance, if $t=2$ the two nodes can be disconnected in slot $s-1$, connected in slots $s$ and $s+1$, and disconnected in slot $s+2$, where $s = 2, \ldots, \tau-2$. Therefore, the percentage of observation time where a duration of $t$ slots can be realized is 
\begin{equation}
\label{eq:g_tau}
g_\tau(t) \coloneqq \frac{\tau-t-1}{\tau}.
\end{equation}
Clearly, for any finite $t$, $g_\tau(t) \to 1$ for $\tau \to \infty$. 

For ease of exposition we use the symbol
\begin{equation}
\xi \coloneqq 1-\omega,
\end{equation}
and observe the following:
\begin{enumerate}
\item [(i)] The probability that two nodes $i$ and $j$ are not connected in a slot $s$ is $1-p_{ij}$, where $p_{ij}$ is given by~(\ref{eq:p_s1}).
\item [(ii)] The probability that $i$ and $j$ are connected in slot $s+1$, given that they are not connected in slot $s$, is $\xi p_{ij}$.
\item [(iii)]  The probability that $i$ and $j$ are connected in slots $s+2, \ldots, t$ given that they are connected in slot $s+1$, is $[1-\xi(1-p_{ij})]^{t-1}$.
\item [(iv)]  The probability that $i$ and $j$ are not connected in slot $t+1$, given that they are connected in slot $t$, is $\xi(1-p_{ij})$.
\end{enumerate} 
It is easy to see that the probability $r_\textnormal{c}(t ; \kappa_i, \kappa_j, \Delta\theta_{ij})$ is the product of $g_\tau(t)$ and the probabilities in points (i) to (iv) above, 
\begin{equation}
\label{eq:p_c}
r_\textnormal{c}(t; \kappa_i, \kappa_j, \Delta\theta_{ij}) = g_\tau(t) \xi^2 p_{ij} (1-p_{ij})^2 [1-\xi(1- p_{ij})]^{t-1}.
\end{equation}
We note that we do not consider the cases when the first (last) of the $t$ slots in which two nodes can be connected starts (ends) at the beginning (end) of the observation period $\tau$. To account for these cases, one needs to add the extra term $(2/\tau) \xi p_{ij} (1-p_{ij}) [1-\xi(1- p_{ij})]^{t-1}$ on the right hand side of~(\ref{eq:p_c}). This term becomes insignificant for any finite $t$ as $\tau \to \infty$. 

The contact distribution, $P_{\textnormal{c}}(t)$, gives the probability that two nodes connect for exactly $t$ consecutive slots, given that they connect, i.e., given that $t \geq 1$. We can write
\begin{align}
\label{eq:p_c_norm}
P_{\textnormal{c}}(t) =\frac{r_\textnormal{c}(t)}{\sum_{j=1}^{\tau-2} r_\textnormal{c}(j)} \propto r_\textnormal{c}(t).
\end{align}
In the above relation, $r_\textnormal{c}(t)$ is obtained by removing the condition on $\kappa_i, \kappa_j$ and $\Delta \theta_{ij}$ from~(\ref{eq:p_c}),
\begin{equation}
r_\textnormal{c}(t) = \int \int \int r_\textnormal{c}(t ; \kappa, \kappa', \Delta\theta) \rho(\kappa) \rho(\kappa') f(\Delta \theta)  \mathrm{d} \kappa \mathrm{d} \kappa' \mathrm{d} \Delta\theta,
\end{equation}
where $\rho(\kappa)$ is the PDF of $\kappa$, while $f(\Delta \theta)=1/\pi$ is the PDF of $\Delta \theta$. 

We note that empirically $P_\textnormal{c}(t)$ is computed as described in the caption of Fig.~\ref{figAll}. Specifically, given a set of (nonzero) contact durations, the empirical $P_\textnormal{c}(t)$ is given by the ratio $n_t/\sum_j n_j$, where $n_t$ is the number of contact durations in the set that have length $t$.

Removing the condition on $\Delta\theta_{ij}$ from~(\ref{eq:p_c}) yields
\begin{align}
\label{eq:full_contact_integral} 
\nonumber r_\textnormal{c}(t ; \kappa_i, \kappa_j ) &=  \frac{1}{\pi} \int \limits_0^\pi  r_\textnormal{c}(t ; \kappa_i, \kappa_j, \Delta\theta) \mathrm{d} \Delta\theta\\
\nonumber &= g_\tau(t) \frac {2 \mu \kappa_i \kappa_j T \xi^2} {N}\\
 & \times \int \limits_{u_{ij}^{\textnormal{min}}}^{1} u^{-T}  (1 - u)^{1+T} [1-\xi (1-u)] ^{t-1} \mathrm{d}u, 
\end{align}
where
\begin{equation}
\label{eq:uij_min}
u_{ij}^{\textnormal{min}} \coloneqq \frac{1}{1+(\frac{N}{2 \mu \kappa_i \kappa_j})^{\frac{1}{T}}}.
\end{equation}
To reach~(\ref{eq:full_contact_integral}), we perform the change of integration variable $u \coloneqq 1/[1+(\frac{N \Delta \theta}{2 \pi \mu \kappa_i \kappa_j})^{1/T}]$. 

\begin{figure}
\includegraphics[width=3.2in]{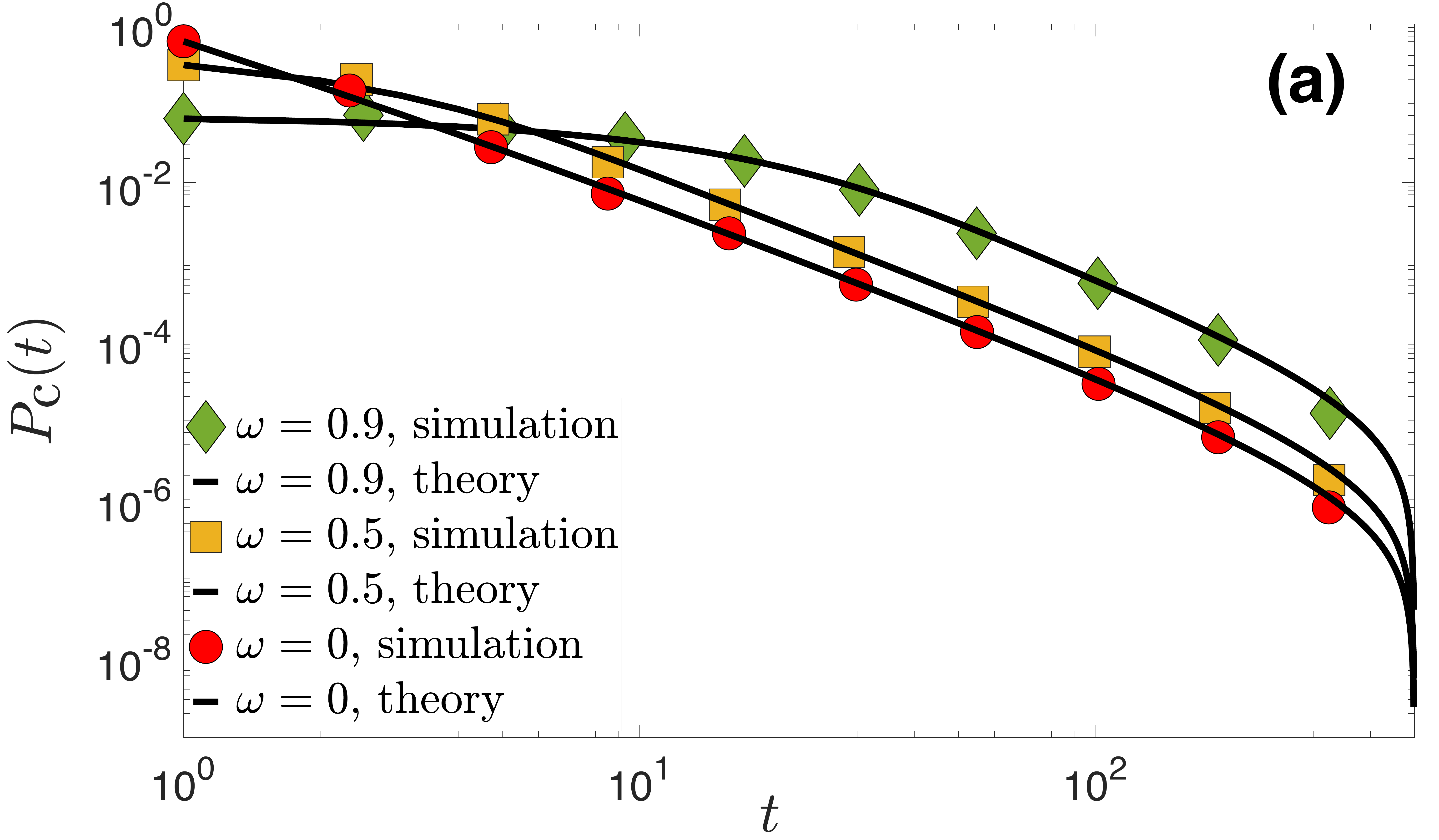}
\includegraphics[width=3.2in]{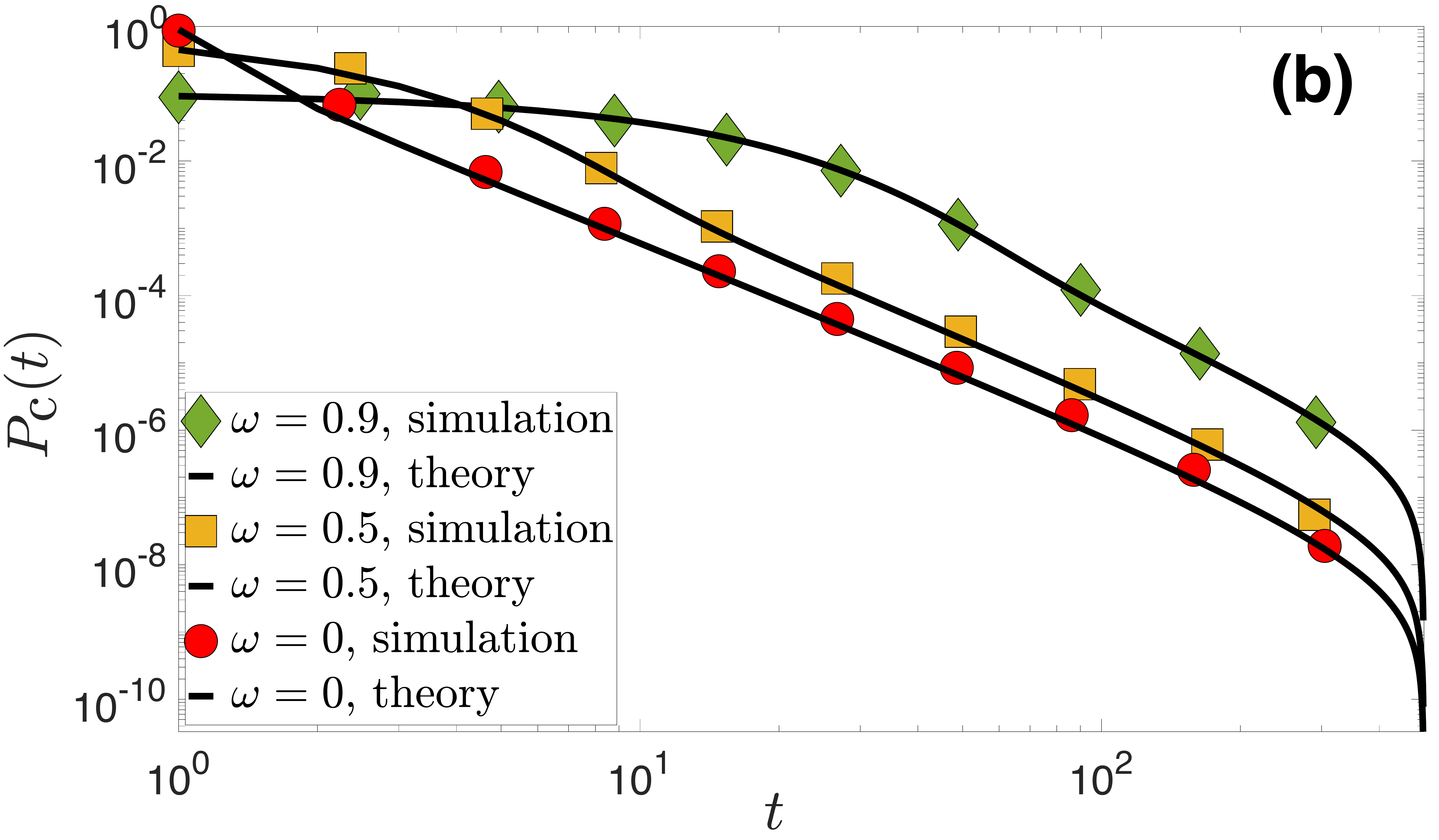}
\caption{Contact distribution in simulated networks with the $\omega$-dynamic-$\mathbb{S}^1$~vs.~theoretical predictions given by~(\ref{eq:p_c_norm}), with $r_\textnormal{c}(t)$ as in~(\ref{eq:approx1}). The number of nodes is $N=500$, the average node degree is $\bar{k}=5$, $\tau=500$, and $\rho(\kappa)=\delta(\kappa-\bar{k})$, where $\delta$ is the Dirac delta function. In (a) $T=0.2$, and in (b) $T=0.8$. Results are shown for different values of the link-persistence probability $\omega=1-\xi$, indicated in the legends. The simulation results are averages over 10 runs and the empirical distributions have been binned logarithmically. The theoretical predictions are given by the solid lines. All axes are in logarithmic scale. 
\label{fig:val1}}
\end{figure}

For $N \to \infty$, $u_{ij}^{\textnormal{min}} \to 0$, and from~(\ref{eq:full_contact_integral}) we have the following limit:
\begin{align}
\label{eq:lim1} 
\nonumber \lim_{N \to \infty} N r_\textnormal{c}(t ; \kappa_i, \kappa_j ) = g_\tau(t) 2 \mu \kappa_i \kappa_j T \xi^2\\
\times \int \limits_0^1 u^{-T}  (1 - u)^{1+T} [1-\xi (1-u)] ^{t-1} \mathrm{d}u.
\end{align}
Removing now the condition on $\kappa_i$ and $\kappa_j$ gives
\begin{align}
\label{eq:lim1_uncon}
\nonumber \lim_{N \to \infty} N r_\textnormal{c}(t) = \lim_{N \to \infty} N \int \int r_\textnormal{c}(t; \kappa, \kappa') \rho(\kappa) \rho(\kappa') \mathrm{d} \kappa \mathrm{d} \kappa'\\
\nonumber = \int \int \lim_{N \to \infty} N r_\textnormal{c}(t; \kappa, \kappa') \rho(\kappa) \rho(\kappa') \mathrm{d} \kappa \mathrm{d} \kappa'\\
= g_\tau(t) 2 \mu \bar{\kappa}^2 T \xi^2 \int \limits_0^1 u^{-T}  (1 - u)^{1+T} [1-\xi (1-u)] ^{t-1} \mathrm{d}u.
\end{align}
We note that we can exchange the order of the limit with the integrals in~(\ref{eq:lim1_uncon}) since $\int \kappa \rho(\kappa) \mathrm{d} \kappa=\bar{\kappa} < \infty$. Further, we note that~(\ref{eq:lim1_uncon}) holds irrespective of the form of $\rho(\kappa)$. Substituting $\mu$ with its expression in~(\ref{eq:mu}), and evaluating the integral in (\ref{eq:lim1_uncon}) (see Appendix~\ref{sec:Appendix1}), yields 
\begin{equation}
\label{eq:lim1_simpler}
\lim_{N \to \infty} N r_\textnormal{c}(t) = g_\tau(t) \frac{\bar{k} T (1+T) \xi^2}{2} {}_2 F_{1} [2+T, 1-t; 3; \xi],
\end{equation}
where ${}_2 F_1[a, b; c; z]$ is the Gauss hypergeometric function~\cite{special_functions_book}.
Therefore, for sufficiently large $N$ we can write
\begin{equation}
\label{eq:approx1}
r_\textnormal{c}(t) \approx g_\tau(t) \frac{\bar{k} T (1+T) \xi^2}{2N} {}_2 F_{1} [2+T, 1-t; 3; \xi].
\end{equation}
Figure~\ref{fig:val1} validates the above analysis.

\begin{figure}
\includegraphics[width=3.1in]{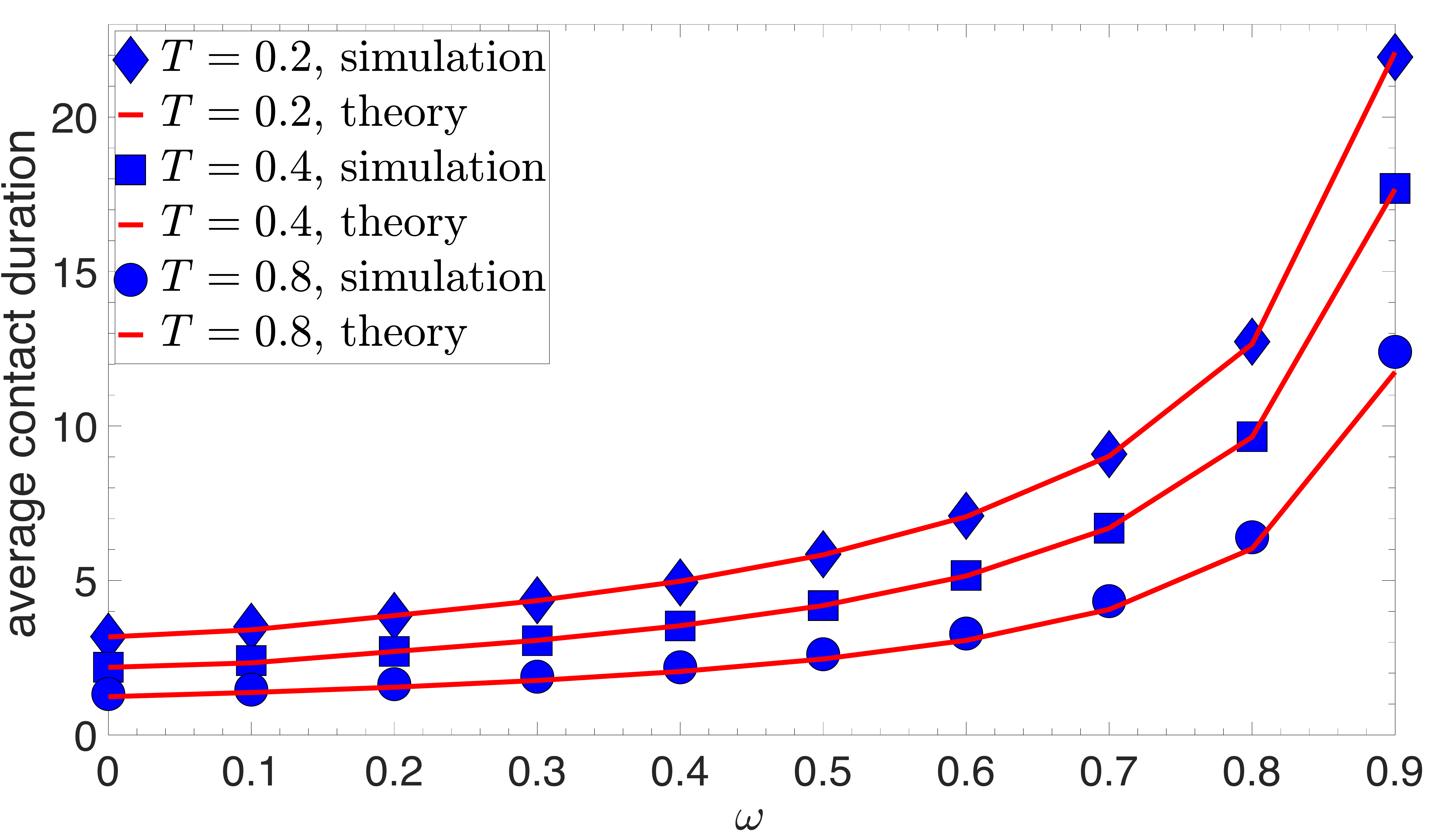}
\caption{Average contact duration $\bar{t}_\textnormal{c}$ as a function of the link-persistence probability $\omega$. Results are shown for different values of the temperature $T$ indicated in the legend. All other simulation parameters are the same as in Fig.~\ref{fig:val1}. The red solid lines show the theoretical predictions given by $\bar{t}_\textnormal{c}=\sum_{t=1}^{\tau-2} t P_\textnormal{c}(t)$.}
\label{fig:ave_contact}
\end{figure}

We note that the average contact duration, $\bar{t}_\textnormal{c}=\sum_{t=1}^{\tau-2} t P_\textnormal{c}(t)$, depends on both the temperature $T$ and the link-persistence probability $\omega=1-\xi$, as dictated by~(\ref{eq:approx1}). In particular, $\bar{t}_\textnormal{c}$ increases with decreasing $T$ or with increasing $\omega$; see Fig.~\ref{fig:ave_contact}.

For $\xi \to 1$, $r_\textnormal{c}(t)$ becomes the one in the dynamic-$\mathbb{S}^{1}$ model~\cite{Papadopoulos2019},
\begin{equation}
\label{eq:cw0}
r_\textnormal{c}(t) \approx g_\tau(t) \frac{\bar{k} T (1+T)}{N \Gamma{(1-T)}} \frac{ \Gamma{(t-T)}}{\Gamma{(t+2)}}.
\end{equation}
For $t \gg 1$, $\Gamma{(t-T)}/\Gamma{(t+2)} \approx 1/t^{2+T}$, while for $t \ll \tau$, $g_\tau(t) \approx 1$. Therefore, for $1 \ll t \ll \tau$,~(\ref{eq:cw0}) decays as a power law,
\begin{equation}
r_\textnormal{c}(t) \propto \frac{1}{t^{2+T}}.
\end{equation}
Interestingly, below we show that for sufficiently large $t$, $r_\textnormal{c}(t)$ also decays as the above power law for all $\xi \in (0, 1)$.

\begin{figure}
\includegraphics[width=3.2in]{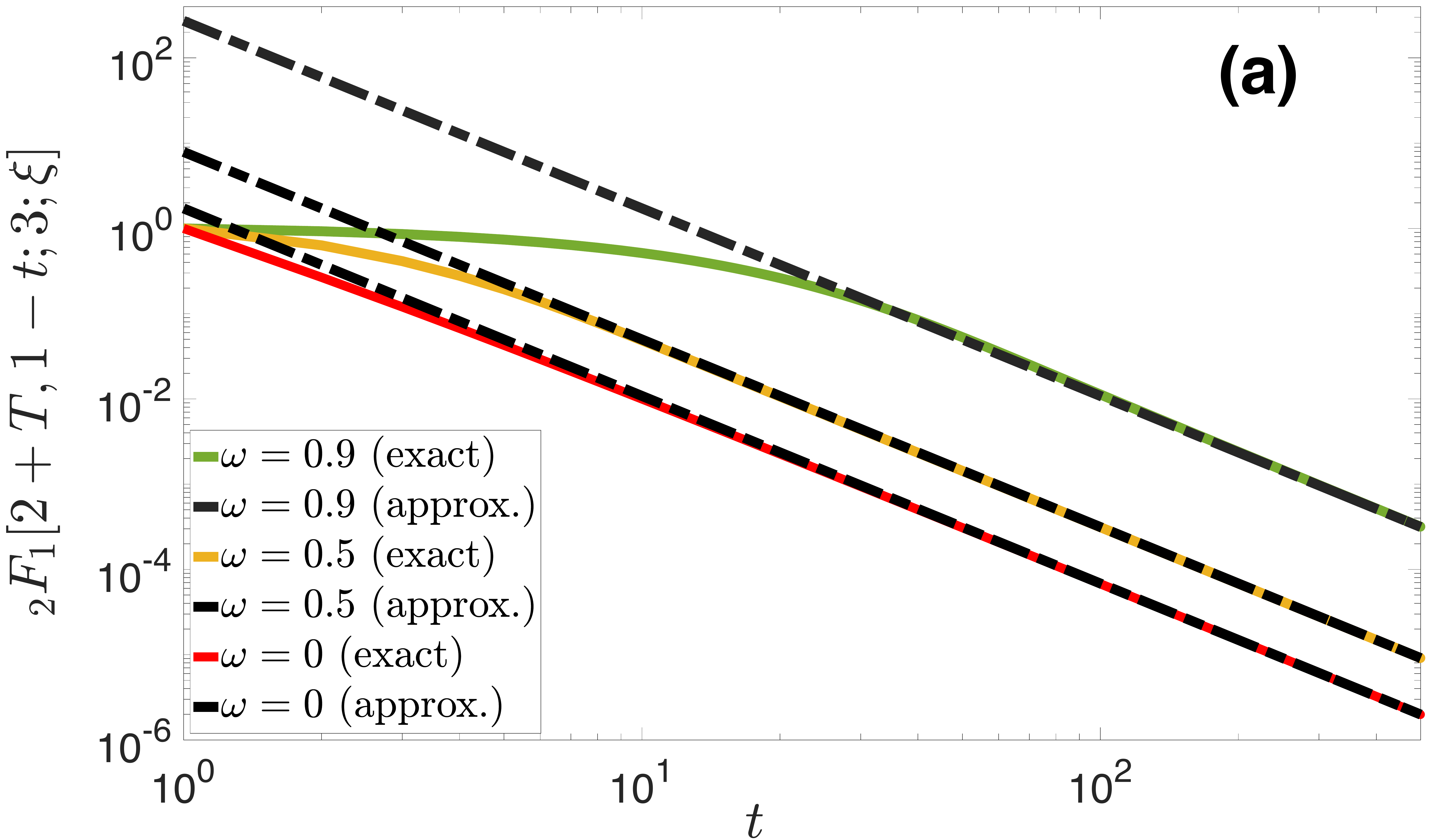}
\includegraphics[width=3.2in]{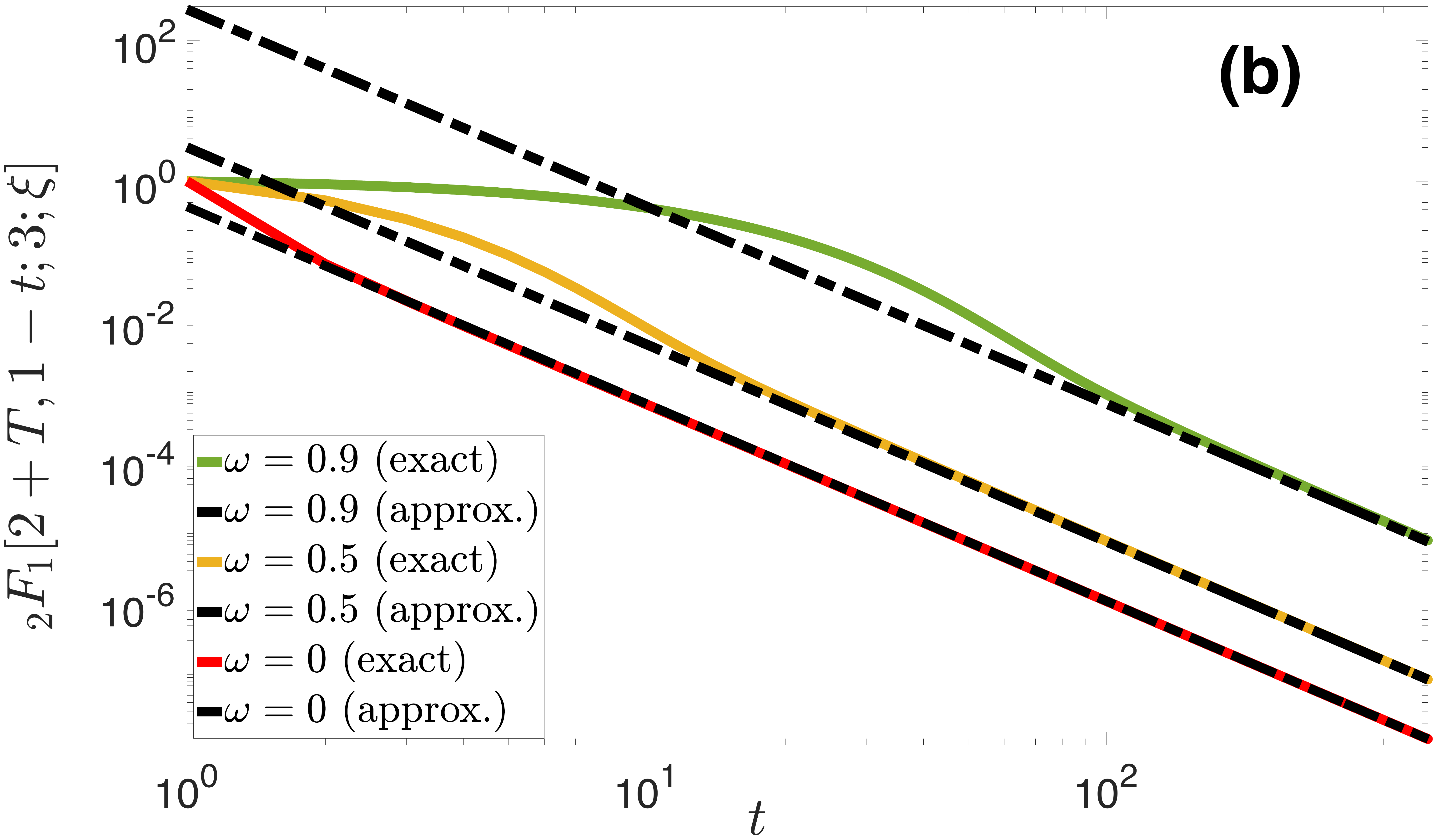}
\caption{ ${}_2 F_{1} [2+T,1-t; 3; \xi]$~vs.~the approximation for large $t$ in~(\ref{eq:approxT2}). In (a)~$T=0.2$, and in (b)~$T=0.8$. Results are shown for different values of $\omega=1-\xi$ indicated in the legends. The solid lines are the exact results, while the dashed-dotted lines are the corresponding approximations given by~(\ref{eq:approxT2}). All axes are in logarithmic scale.
\label{fig:approx1_val}}
\end{figure}

\textbf{Tail of $r_\textnormal{c}(t)$.} To deduce the behavior of the tail of $r_\textnormal{c}(t)$ for $\xi \in (0, 1)$, we utilize an asymptotic expansion for the hypergeometric function ${}_2 F_{1} [a, b; c; z]$ for $|b| \to \infty$, given in section~2.3.2 of Ref.~\cite{bateman1953higher} (Eq.~(15) on p.~77). This expansion allows us to express the hypergeometric function in~(\ref{eq:approx1}) for $t\to \infty$, as

\begin{align}
\label{eq:T1}
\nonumber  {}_2 F_{1} [2+T, 1-t; 3; \xi] & = \Bigg\{\frac{2 (-1)^{2+T}}{\Gamma(1-T)} [\xi(1-t)]^{-(2+T)}\\
 \nonumber & + \frac{2 e^{-\xi (t-1)}}{\Gamma(2+T) }  [\xi(1-t)]^{-(1-T)}\Bigg\}\\
 &\times [1+O(|\xi (1-t)|^{-1})]. 
\end{align} 
Equation~(\ref{eq:T1}) means that for sufficiently large $\xi t $ we can write
\begin{align}
\label{eq:approxT1}
\nonumber  {}_2 F_{1} [2+T, 1-t; 3; \xi] & \approx \frac{2}{\Gamma(1-T)} \frac{1}{(\xi t)^{2+T}}\\
& - \frac{2 (-1)^T}{\Gamma(2+T)}\frac{1}{e^{\xi t} (\xi t)^{1-T}}.
\end{align}
Further, since the dominant term in~(\ref{eq:approxT1}) is the first for large $\xi t$, we can write the following simplified expression:         
\begin{equation}
\label{eq:approxT2}
 {}_2 F_{1} [2+T,1-t; 3; \xi] \approx \frac{2}{\Gamma(1-T)} \frac{1}{(\xi t)^{2+T}} \propto \frac{1}{t^{2+T}}.
\end{equation}
We note that since $\xi$ is fixed, $\xi \in (0, 1)$, the approximations in~(\ref{eq:approxT1}) and~(\ref{eq:approxT2}) come into effect for sufficiently large $t$. Figure~(\ref{fig:approx1_val}) validates the above analysis.

Therefore, for large $t \ll \tau$, $r_\textnormal{c}(t)$ in~(\ref{eq:approx1}) is proportional to $1/t^{2+T}$ for all $\xi \in (0, 1]$. Next, we turn our attention to the intercontact distribution. 

\subsection{Intercontact distribution}

To analyze the intercontact distribution we follow a similar procedure as in the contact distribution. Let $r_\textnormal{ic}(t; \kappa_i, \kappa_j, \Delta\theta_{ij})$ be the probability to observe a sequence of exactly $t$ consecutive slots, where two nodes  $i$ and $j$ with latent degrees $\kappa_i$ and $\kappa_j$ and angular distance $\Delta \theta_{ij}$ are not connected, $t=1, 2, \ldots, \tau-2$. This probability is the percentage of observation time $\tau$ where we observe a slot where these two nodes are connected, followed by $t$ slots where they are not connected, followed by a slot where they are again connected.  

We observe the following:
\begin{enumerate}
\item [(i)] The probability that two nodes $i$ and $j$ are connected in a slot $s$ is $p_{ij}$, given by~(\ref{eq:p_s1}).
\item [(ii)] The probability that $i$ and $j$ are not connected in slot $s+1$, given that they are connected in slot $s$, is $\xi (1-p_{ij})$.
\item [(iii)]  The probability that $i$ and $j$ are not connected in slots $s+2, \ldots, t$ given that they are not connected in slot $s+1$, is $(1-\xi p_{ij})^{t-1}$.
\item [(iv)]  The probability that $i$ and $j$ are connected in slot $t+1$, given that they are not connected in slot $t$, is $\xi p_{ij}$.
\end{enumerate} 
The probability $r_\textnormal{ic}(t ; \kappa_i, \kappa_j, \Delta\theta_{ij})$ is the product of $g_\tau(t)$ in~(\ref{eq:g_tau}) and the probabilities in points (i) to (iv) above, 
\begin{equation}
\label{eq:p_ic}
r_\textnormal{ic}(t; \kappa_i, \kappa_j, \Delta\theta_{ij})= g_\tau(t) \xi^2 p_{ij}^2 (1-p_{ij}) (1-\xi p_{ij})^{t-1}.
\end{equation}
We note that considering adding an extra term on the right hand side of (\ref{eq:p_ic}) analogous to the one discussed below~(\ref{eq:p_c}) would be unnatural here, since by its name an intercontact duration must be enclosed between two contacts.

The intercontact distribution, $P_{\textnormal{ic}}(t)$, gives the probability that two nodes disconnect for exactly $t$ consecutive slots, given that they disconnect, i.e., given that $t \geq 1$. We can write
\begin{align}
\label{eq:p_ic_norm}
P_{\textnormal{ic}}(t) =\frac{r_\textnormal{ic}(t)}{\sum_{j=1}^{\tau-2} r_\textnormal{ic}(j)} \propto r_\textnormal{ic}(t).
\end{align}
In the above relation, $r_\textnormal{ic}(t)$ is obtained by removing the condition on $\kappa_i, \kappa_j$ and $\Delta \theta_{ij}$ from~(\ref{eq:p_ic}),
\begin{equation}
\label{eq:full_intercontact_integral1}
r_\textnormal{ic}(t) = \int \int \int r_\textnormal{ic}(t ; \kappa, \kappa', \Delta\theta) \rho(\kappa) \rho(\kappa') f(\Delta \theta)  \mathrm{d} \kappa \mathrm{d} \kappa' \mathrm{d} \Delta\theta.
\end{equation}
We note that as with $P_\textnormal{c}(t)$, given a set of (nonzero) intercontact durations, the empirical $P_\textnormal{ic}(t)$ is given by the ratio $n_t/\sum_j n_j$, where $n_t$ is the number of intercontact durations in the set that have length $t$.

Removing the condition on $\Delta\theta_{ij}$ from~(\ref{eq:full_intercontact_integral1}) yields
\begin{align}
\label{eq:full_intercontact_integral} 
\nonumber r_\textnormal{ic}(t ; \kappa_i, \kappa_j ) &=  \frac{1}{\pi} \int \limits_0^\pi  r_\textnormal{ic}(t ; \kappa_i, \kappa_j, \Delta\theta) \mathrm{d} \Delta\theta\\
\nonumber &= g_\tau(t)\frac{2 \mu \kappa_i \kappa_j T \xi^2} {N}\\
 & \times \int \limits_{u_{ij}^{\textnormal{min}}}^{1} u^{1-T}  (1 - u)^T (1-\xi u)^{t-1} \mathrm{d}u, 
\end{align}
where
$u_{ij}^{\textnormal{min}}$ is given by~(\ref{eq:uij_min}). To reach~(\ref{eq:full_intercontact_integral}), we again perform the change of integration variable $u \coloneqq 1/[1+(\frac{N \Delta \theta}{2 \pi \mu \kappa_i \kappa_j})^{1/T}]$.

\begin{figure}
\includegraphics[width=3.2in]{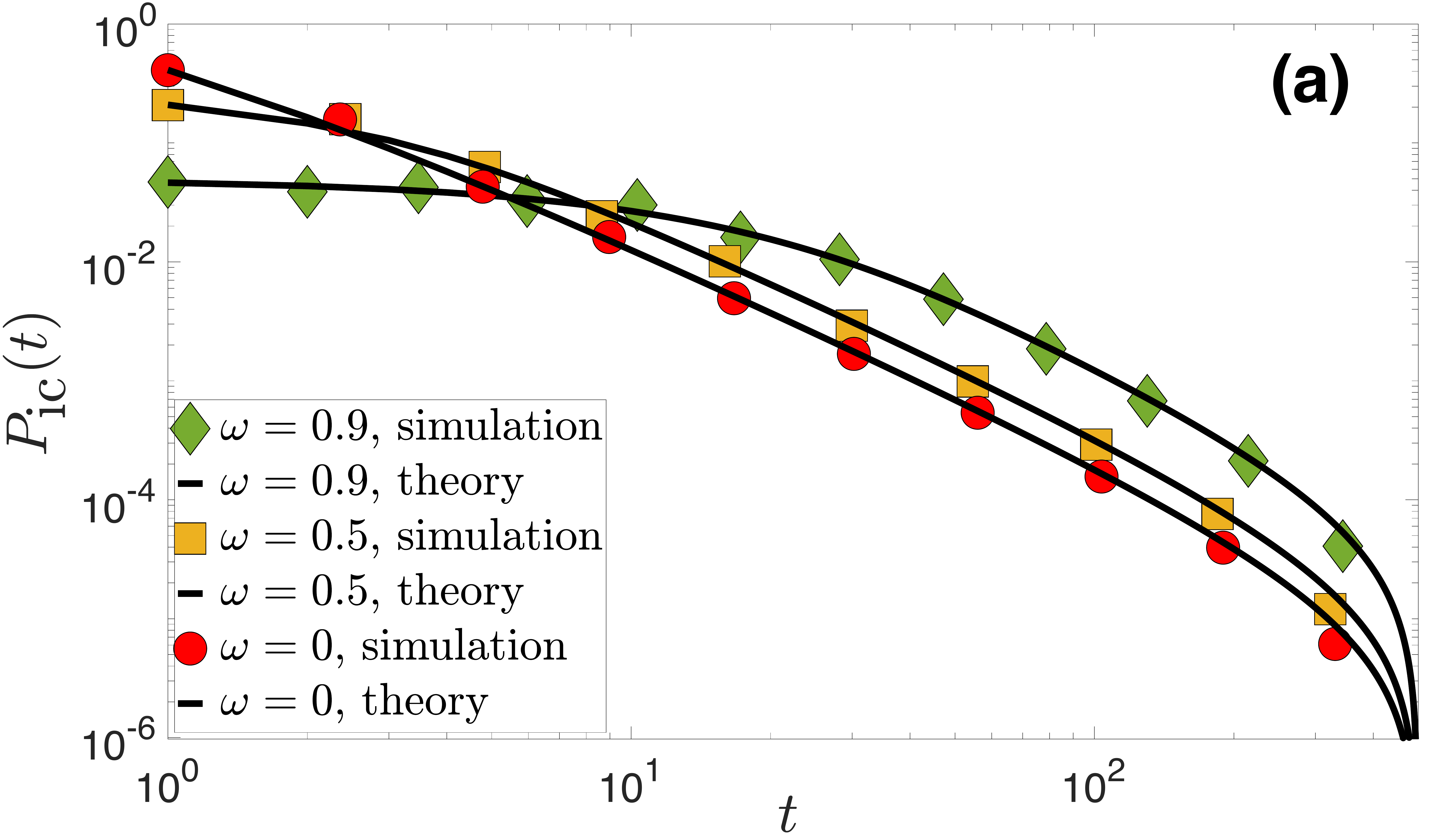}
\includegraphics[width=3.2in]{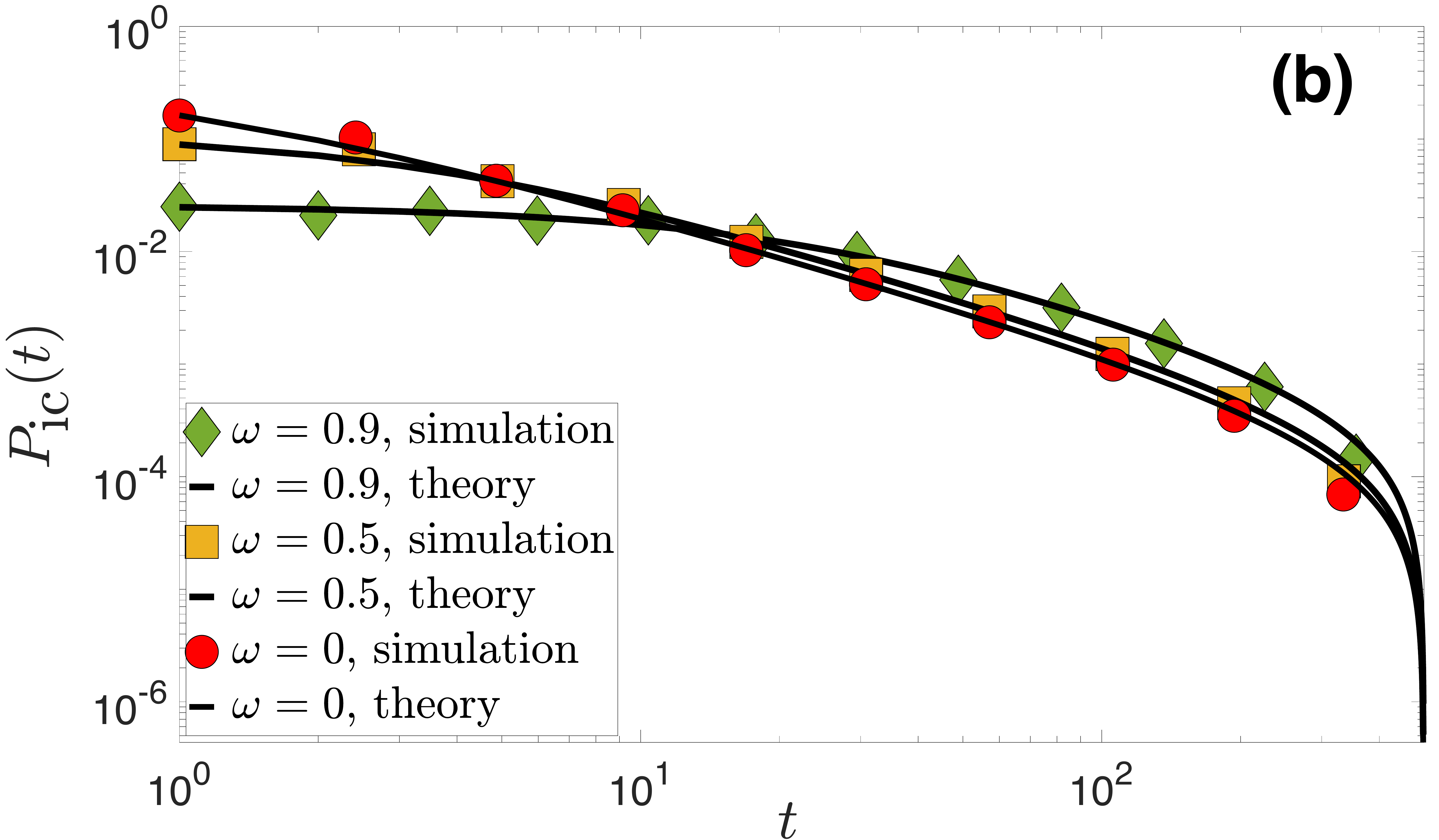}
\caption{Same as in Fig.~\ref{fig:val1} but for the intercontact distribution. The theoretical predictions are given by~(\ref{eq:p_ic_norm}), with $r_\textnormal{ic}(t)$ as in~(\ref{eq:approx2}). In~(a) $T=0.2$, and in~(b) $T=0.8$.   
\label{fig:val2}}
\end{figure}

For $N \to \infty$, $u_{ij}^{\textnormal{min}} \to 0$, and from~(\ref{eq:full_intercontact_integral}) we have the following limit:
\begin{align}
\label{eq:lim2} 
\nonumber \lim_{N \to \infty} N r_\textnormal{ic}(t ; \kappa_i, \kappa_j ) = g_\tau(t) 2 \mu \kappa_i \kappa_j T \xi^2\\
\times \int \limits_0^1 u^{1-T}  (1 - u)^T (1-\xi u)^{t-1} \mathrm{d}u.
\end{align}
We can now compute
\begin{align}
\label{eq:lim2_uncon}
\nonumber \lim_{N \to \infty} N r_\textnormal{ic}(t) = \lim_{N \to \infty} N \int \int r_\textnormal{ic}(t; \kappa, \kappa') \rho(\kappa) \rho(\kappa') \mathrm{d} \kappa \mathrm{d} \kappa'\\
\nonumber = \int \int \lim_{N \to \infty} N r_\textnormal{ic}(t; \kappa, \kappa') \rho(\kappa) \rho(\kappa') \mathrm{d} \kappa \mathrm{d} \kappa'\\
= g_\tau(t) 2 \mu \bar{\kappa}^2 T \xi^2 \int \limits_0^1 u^{1-T}  (1 - u)^T (1-\xi u)^{t-1} \mathrm{d}u.
\end{align}
Substituting $\mu$ with its expression in~(\ref{eq:mu}), and evaluating the integral in (\ref{eq:lim2_uncon}) (see Appendix~\ref{sec:Appendix2}), yields 
\begin{equation}
\label{eq:lim2_simpler}
\lim_{N \to \infty} N r_\textnormal{ic}(t) = g_\tau(t) \frac{\bar{k} T (1-T) \xi^2}{2} {}_2 F_{1} [2-T, 1-t; 3; \xi].
\end{equation}
Therefore, for sufficiently large $N$ we can write
\begin{equation}
\label{eq:approx2}
r_\textnormal{ic}(t) \approx g_\tau(t) \frac{\bar{k} T (1-T) \xi^2}{2N} {}_2 F_{1} [2-T, 1-t; 3; \xi].
\end{equation}
Figure~\ref{fig:val2} validates the above analysis. 

\begin{figure}
\includegraphics[width=3.1in]{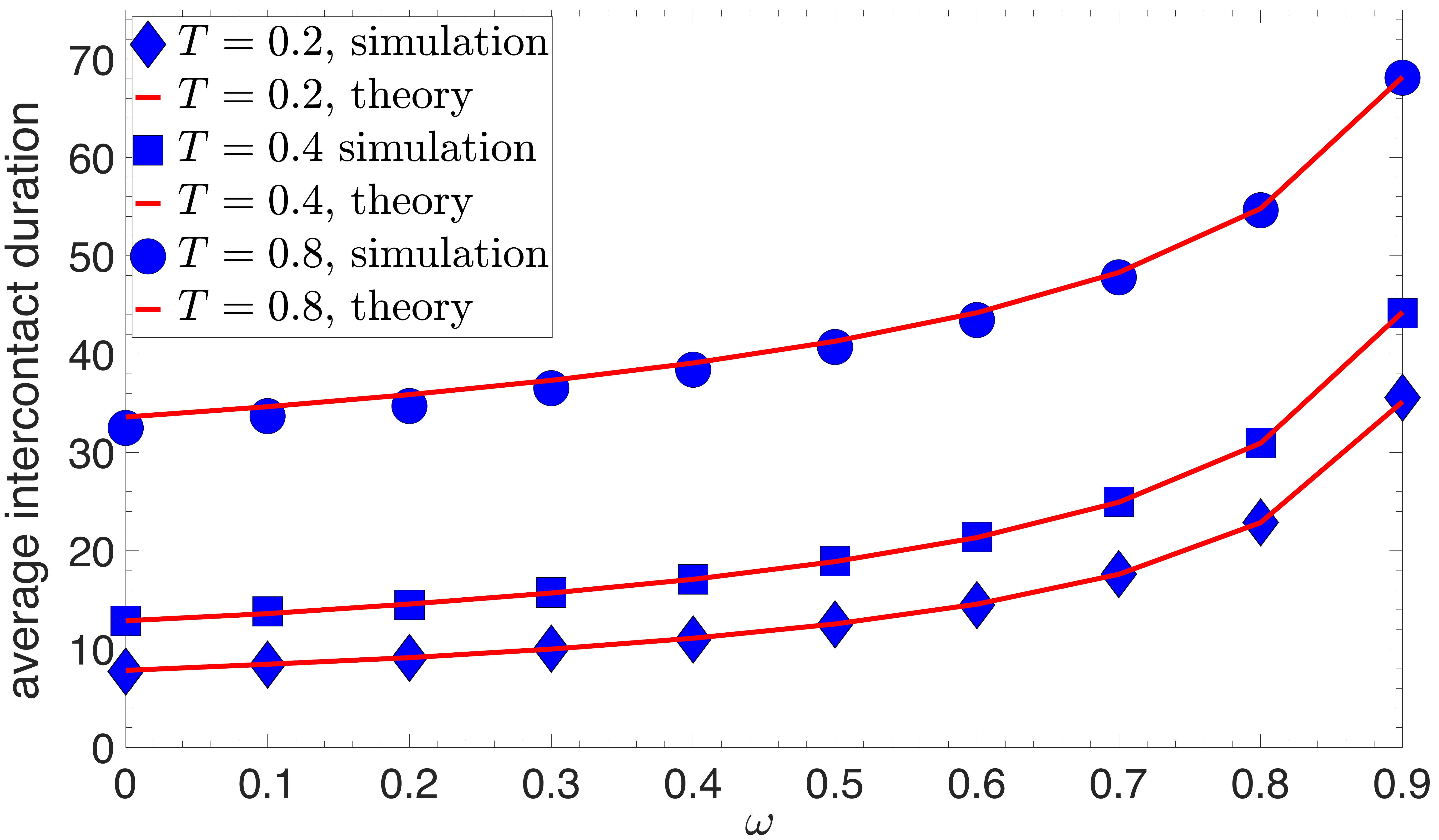}
\caption{Same as in Fig.~\ref{fig:ave_contact} but for the average intercontact duration $\bar{t}_\textnormal{ic}$. The theoretical predictions are given by $\bar{t}_\textnormal{ic}=\sum_{t=1}^{\tau-2} t P_\textnormal{ic}(t)$.}
\label{fig:ave_intercontact}
\end{figure}

The average intercontact duration, $\bar{t}_\textnormal{ic}=\sum_{t=1}^{\tau-2} t P_\textnormal{ic}(t)$, depends on both the temperature $T$ and the link-persistence probability $\omega=1-\xi$, as dictated by~(\ref{eq:approx2}). In particular, $\bar{t}_\textnormal{ic}$ increases with increasing $T$ or with increasing $\omega$; see Fig.~\ref{fig:ave_intercontact}.

For $\xi \to 1$, $r_\textnormal{ic}(t)$ becomes the one in the dynamic-$\mathbb{S}^{1}$ model~\cite{Papadopoulos2019},
\begin{equation}
\label{eq:icw0}
r_\textnormal{ic}(t) \approx g_\tau(t) \frac{\bar{k} T (1-T)}{N \Gamma{(1+T)}} \frac{\Gamma{(t+T)}}{\Gamma{(t+2)}}.
\end{equation}
For $t \gg 1$, $\Gamma{(t+T)}/\Gamma{(t+2)} \approx 1/t^{2-T}$, while for $t \ll \tau$, $g_\tau(t) \approx 1$. Therefore, for $1 \ll t \ll \tau$,~(\ref{eq:icw0}) decays as a power law,
\begin{equation}
r_\textnormal{ic}(t) \propto \frac{1}{t^{2-T}}.
\end{equation}
Below, we show that for sufficiently large $t$, $r_\textnormal{ic}(t)$ also decays as the above power law for all $\xi \in (0, 1)$.

\begin{figure}
\includegraphics[width=3.2in]{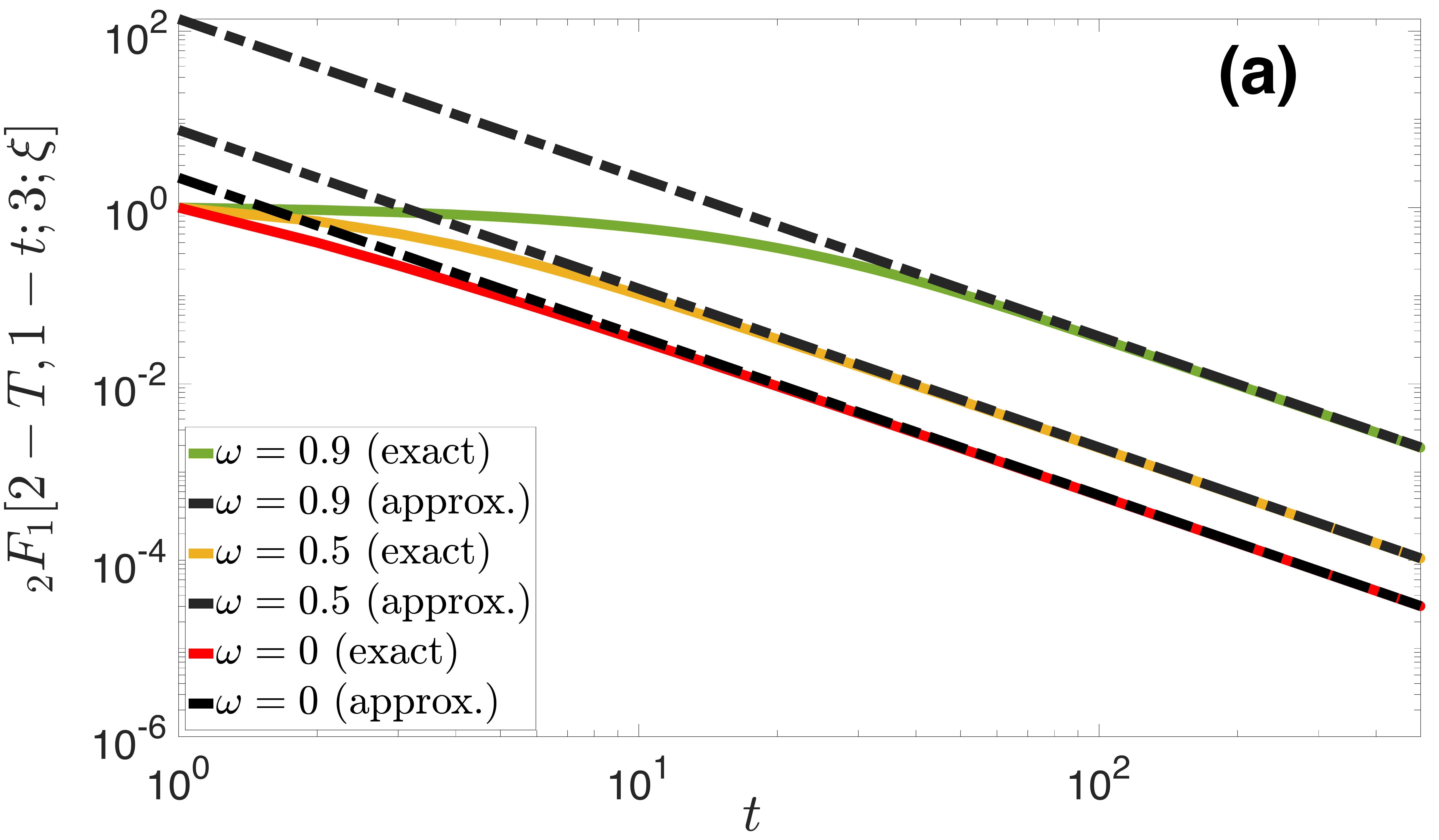}
\includegraphics[width=3.2in]{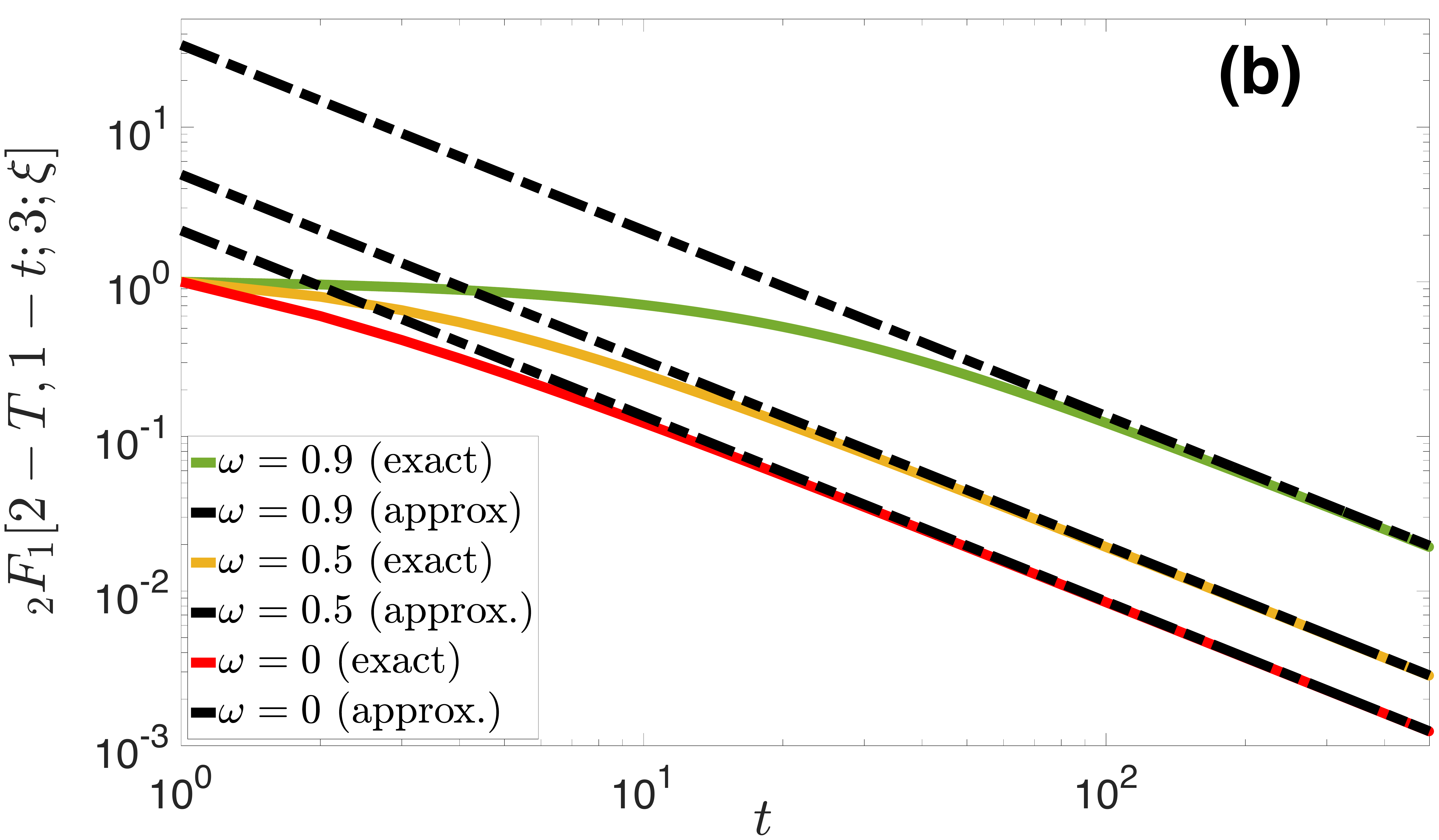}
\caption{ ${}_2 F_{1} [2-T,1-t; 3; \xi]$~vs.~the approximation for large $t$ in~(\ref{eq:approxT4}). In (a)~$T=0.2$, and in (b)~$T=0.8$. Results are shown for different values of $\omega=1-\xi$. The solid lines are the exact results, while the dashed-dotted lines are the corresponding approximations given by~(\ref{eq:approxT4}).
\label{fig:approx2_val}}
\end{figure}

\textbf{Tail of $r_\textnormal{ic}(t)$.} To deduce the behavior of the tail of $r_\textnormal{ic}(t)$ for $\xi \in (0, 1)$, we utilize again the expansion for ${}_2 F_{1} [a, b; c; z]$ for $|b| \to \infty$ given in Eq.~(15) on p.~77 of Ref.~\cite{bateman1953higher}. Using this expansion, we can express the hypergeometric function in~(\ref{eq:approx2}) for $t \to \infty$ as

\begin{align}
\label{eq:T2}
\nonumber  {}_2 F_{1} [2-T, 1-t; 3; \xi] & = \Bigg\{\frac{2 (-1)^{2-T}}{\Gamma(1+T)} [\xi(1-t)]^{-(2-T)}\\
 \nonumber & + \frac{2 e^{-\xi (t-1)}}{\Gamma(2-T) }  [\xi(1-t)]^{-(1+T)}\Bigg\}\\
 &\times [1+O(|\xi (1-t)|^{-1})]. 
\end{align} 
Therefore, for sufficiently large $\xi t$ we can write
\begin{align}
\label{eq:approxT3}
\nonumber  {}_2 F_{1} [2-T, 1-t; 3; \xi] & \approx \frac{2}{\Gamma(1+T)} \frac{1}{(\xi t)^{2-T}}\\
& - \frac{2 (-1)^{-T}}{\Gamma(2-T)}\frac{1}{e^{\xi t} (\xi t)^{1+T}}.
\end{align}
Further, since the dominant term in the above relation is the first for large $\xi t$, we can write the following simplified expression:         
\begin{equation}
\label{eq:approxT4}
{}_2 F_{1} [2-T,1-t; 3; \xi] \approx  \frac{2}{\Gamma(1+T)} \frac{1}{(\xi t)^{2-T}} \propto \frac{1}{t^{2-T}}.
\end{equation}
Since $\xi$ is fixed, $\xi \in (0, 1)$, the approximations in~(\ref{eq:approxT3}) and~(\ref{eq:approxT4}) come into effect for sufficiently large $t$. Figure~\ref{fig:approx2_val} validates the above analysis. 

Therefore, for large $t \ll \tau$, $r_\textnormal{ic}(t)$ in~(\ref{eq:approx2}) is proportional to $1/t^{2-T}$ for all $\xi \in (0, 1]$. 

\subsection{Expected time-aggregated degree}
\label{sec:kaggr}

Here we turn our attention to the expected time-aggregated degree, and show its dependence on both the temperature $T$ and the link-persistence probability $\omega$. The expected time-aggregated node degree can be written as
\begin{equation}
\label{eq:c1}
\bar{k}_\textnormal{aggr}=(N-1)(1-r_0),
\end{equation}
where $1-r_0$ is the probability that two nodes connect at least once during the observation interval $\tau$. Below, we derive the relation for $\bar{k}_\textnormal{aggr}$ for large $N$.

Let $r_0(\kappa_i, \kappa_j, \Delta\theta_{ij})$ be the probability that two nodes $i$ and $j$ with latent degrees $\kappa_i$ and $\kappa_j$ and angular distance $\Delta\theta_{ij}$ do not connect during the observation interval $\tau$. We can write
\begin{equation}
\label{eq:c2}
r_0(\kappa_i, \kappa_j, \Delta\theta_{ij}) = (1-p_{ij}) (1 - \xi p_{ij})^{\tau-1}, 
\end{equation}
where $p_{ij}$ is given by~(\ref{eq:p_s1}). Removing the condition on $\Delta\theta_{ij}$ gives
\begin{align}
\label{eq:c3} 
\nonumber & r_0(\kappa_i, \kappa_j) = \frac{1}{\pi} \int \limits_0^\pi r_0(\kappa_i, \kappa_j, \Delta\theta)\mathrm{d} \Delta\theta\\
\nonumber & =\frac {2\mu\kappa_i\kappa_j T}{N}\int \limits_{u_{ij}^{\textnormal{min}}}^1 u^{-(1+T)}(1-u)^T (1-\xi u)^{\tau-1}\mathrm{d}u\\ 
\nonumber  & =  \frac {2\mu\kappa_i\kappa_j T}{N} \Big\{\frac{(u_{ij}^{\textnormal{min}})^{-T}}{T}F_1[-T, -T, -\tau; 1-T; u_{ij}^{\textnormal{min}}, \xi u_{ij}^{\textnormal{min}}]\\
\nonumber & - \frac{\xi (u_{ij}^{\textnormal{min}})^{1-T}}{1-T} F_1 [1-T, -T, 1-\tau; 2-T; u_{ij}^{\textnormal{min}}, \xi u_{ij}^{\textnormal{min}}]\\
\nonumber & + \frac{\xi T\pi}{\sin{(T \pi)}} {}_2 F_{1} [1-T, 1-\tau; 2; \xi]\\
& - \frac{\pi}{\sin{(T \pi)}} {}_2 F_{1} [-T, -\tau; 1; \xi]\Big\},
\end{align}
where $u_{ij}^{\textnormal{min}}$ is given by~(\ref{eq:uij_min}). In the above relation, $F_1[a, b_1, b_2; c; x, y]$ is the Appell series, which is a generalization of the hypergeometric function for two variables $x, y$~\cite{special_functions_book}. To reach~(\ref{eq:c3}), we perform the change of integration variable $u \coloneqq 1/[1+(\frac{N \Delta \theta}{2 \pi \mu \kappa_i \kappa_j})^{1/T}]$.

We note that for $x \to 0$ and $y \to 0$, $F_1[a, b_1, b_2; c; x, y] \to 1$. Further, $u_{ij}^{\textnormal{min}} \to 0$ for $N\to \infty$. From~(\ref{eq:c3}), we have the following limit:
\begin{align}
\label{eq:c4} 
\nonumber & \lim_{N \to \infty} N[1-r_0(\kappa_i, \kappa_j)]= \frac{2\mu\kappa_i\kappa_j T \pi}{\sin{(T \pi)}} \Big\{{}_2 F_{1} [-T, -\tau; 1; \xi]\\
& - \xi T {}_2 F_{1} [1-T, 1-\tau; 2; \xi]\Big\}.
\end{align}
Removing the condition on $\kappa_i$ and $\kappa_j$ from~(\ref{eq:c4}), and substituting $\mu$ with its expression in~(\ref{eq:mu}), yields
\begin{align}
\label{eq:c5} 
\nonumber \lim_{N \to \infty} N(1-r_0) &= \Big\{{}_2 F_{1} [-T, -\tau; 1; \xi]\\
&-\xi T {}_2 F_{1} [1-T, 1-\tau; 2; \xi]\Big\} \bar{k}.
\end{align}
Therefore, for sufficiently large $N$ we can write
\begin{equation}
\label{eq:c6}
\bar{k}_\textnormal{aggr} \approx \Big\{ {}_2 F_{1} [-T, -\tau; 1; \xi]-\xi T {}_2 F_{1} [1-T, 1-\tau; 2; \xi]\Big\}\bar{k}.
\end{equation}

\begin{figure}
\includegraphics[width=3.2in]{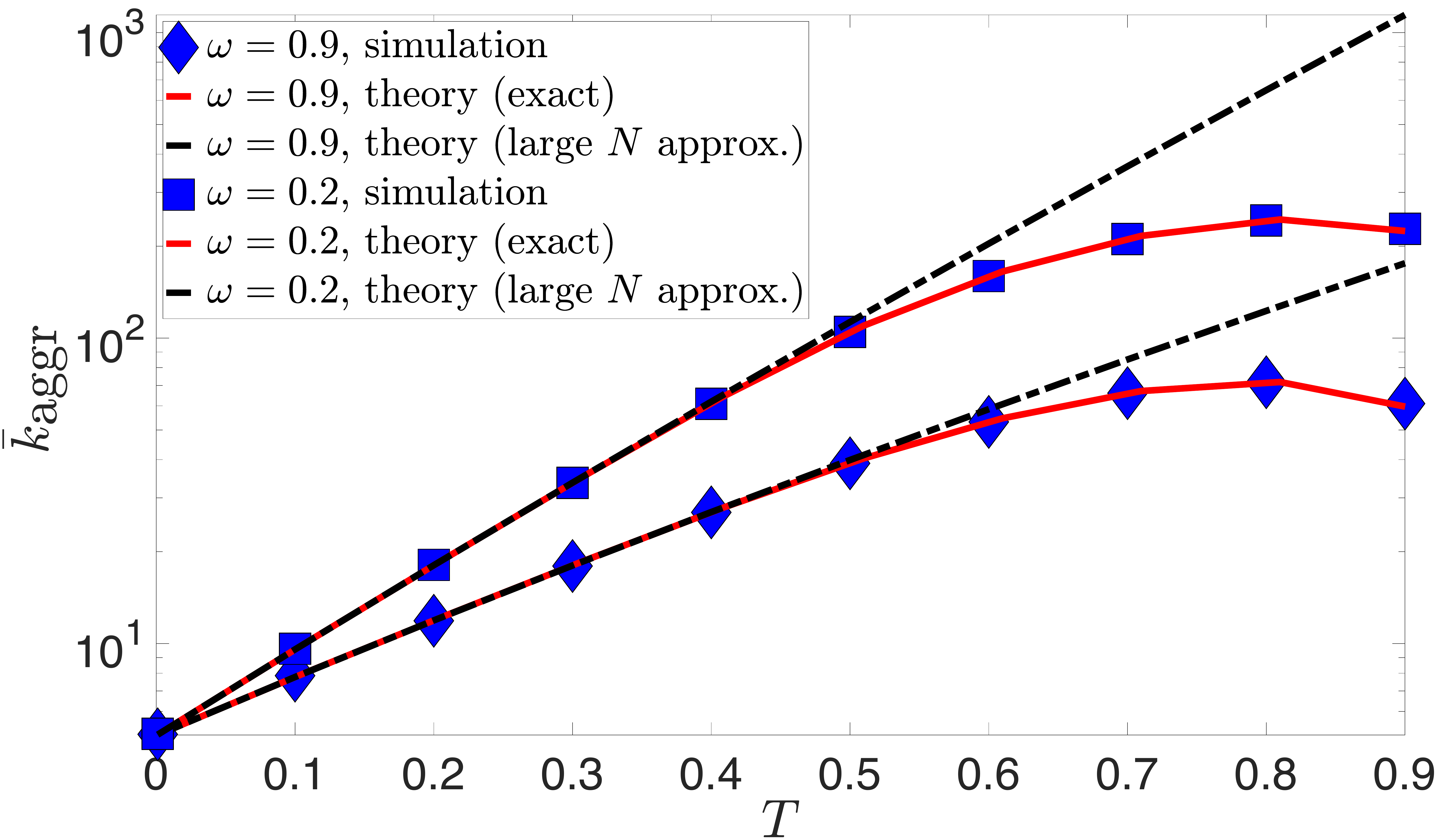}
\caption{Average time-aggregated degree $\bar{k}_\textnormal{aggr}$ as a function of the temperature $T$ in simulated networks~vs.~theoretical predictions. Results are shown for two values of $\omega=1-\xi$ indicated in the legend. All other simulation parameters are the same as in Fig.~\ref{fig:val1}. The red solid lines show the exact theoretical predictions given via (\ref{eq:c1}) and (\ref{eq:c3}). The dashed-dotted lines show the corresponding 
large-$N$ approximations given by~(\ref{eq:c8}). The $y$-axis is in logarithmic scale.}
\label{fig:kaggr_val}
\end{figure}

For $\xi=1$, the above expression becomes the one in the dynamic-$\mathbb{S}^1$ model~\cite{Papadopoulos2019},
\begin{equation}
\label{eq:c7}
\bar{k}_\textnormal{aggr} \approx \frac{\Gamma{(\tau+T)}\bar{k}}{\Gamma{(1+T)}\Gamma{(\tau)}} \approx \frac{\tau^T\bar{k}}{\Gamma{(1+T)}}.
\end{equation}
The last approximation in~(\ref{eq:c7}) holds for $\tau \gg 1$.
Further, we note that we can again utilize the expansion for ${}_2 F_{1} [a, b; c; z]$ for $|b| \to \infty$ given in Eq.~(15) on p.~77 of Ref.~\cite{bateman1953higher} to simplify~(\ref{eq:c6}). Specifically, using this expansion, we can write (details are omitted for brevity) that for sufficiently large $\tau$,
\begin{equation}
\label{eq:c8}
\bar{k}_\textnormal{aggr} \approx \frac{(\xi \tau)^T\bar{k}}{\Gamma{(1+T)}}.
\end{equation}
Fig.~\ref{fig:kaggr_val} validates our analysis. We see that~(\ref{eq:c8}) is a good approximation only for sufficiently low temperatures $T$. In general, to accurately compute the expected time-aggregated degree for any temperature $T$ one would need to remove the condition on $\kappa_i$ and $\kappa_j$ from the exact expression in~(\ref{eq:c3}), a task that could be done numerically for any PDF $\rho(\kappa)$, and use the result in~(\ref{eq:c1}).

\section{Other related work}
\label{sec:related_work}

The work in Ref.~\cite{Perra2012} introduced the activity-driven model (AD), while the work in Ref.~\cite{Perra2017} extended this model to account for node attractiveness. However, the AD is not a geometric network model. Here, we considered a geometric temporal network model based on RHGs, which have been shown to adequately reflect reality. Further, link persistence has not been considered in the context of the AD. Finally, the analysis of the AD has mainly focused on properties of the resulting time-aggregated network, like its degree distribution~\cite{Perra2012}, and not on properties of the resulting temporal network itself, like its (inter)contact distributions. 

The work in Ref.~\cite{Moore2017} proposed temporal extensions of popular static network models (random graphs, configuration model, stochastic block model) and provided algorithms for fitting the proposed extensions to observed network data. Even though this work considers link persistence, it does not consider temporal extensions of geometric network models, nor does it analyze the resulting temporal properties of the proposed extensions in terms of their realism.

The work in Ref.~\cite{mazzarisi2020} considers link persistence (also called stability) in dynamic networks, in conjunction with node hidden variables (or fitnesses) that determine the nodes'
capability of forming links, and it attempts to disentangle the importance of the two mechanisms (link persistence versus node hidden variables) in link formation in the interbank market.  To this end, it considers a  link-persistence model similar to the one we considered here. However, differently from this work, it does not consider RHGs, i.e., networks where the node hidden variables
are their coordinates in their underlying hyperbolic space. Further, it does not analyze emergent dynamical properties, such as the (inter)contact distributions, and the effect of link persistence on them. 

In Ref.~\cite{Papadopoulos2019lp}, a model based on RHGs similar to the $\omega$-dynamic-$\mathbb{S}^1$, but with persistence only for connections (instead for both connections and disconnections), has been shown to better explain the high edge overlap across layers of real multiplex networks, compared to the case in which link persistence is ignored. We also note that the $\omega$-dynamic-$\mathbb{S}^1$ is a special case of the general class of temporal hidden-variable network models considered in Ref.~\cite{hartle2021}, where there are no hidden variable dynamics.
A review of other work related to the concept of persistence in temporal networks and in complex systems in general can be found in Ref.~\cite{persistencereview2022}.

\section{Discussion and conclusion}
\label{sec:conclusion}

We have considered and analyzed a simple dynamical model of RHGs with link persistence, called $\omega$-dynamic-$\mathbb{S}^1$. Despite its simplicity, the model simultaneously reproduces many dynamical properties observed in real systems, while providing flexibility in tuning the average contact and intercontact durations via the link-persistence probability $\omega$. We have analyzed two main properties of interest, i.e., the distributions of contact and intercontact durations, and found that they both decay as power laws in the model with exponents that do not depend on $\omega$. We have also analyzed the expected time-aggregated degree in the model.

In future work, it would be interesting to analyze other temporal network properties, such as the weight and strength distributions, cf. Fig.~\ref{figAll}, and statistics related to components' formation, cf. Ref.~\cite{Papadopoulos2019}. Further, it is desirable to explore generalizations of the model where connections and disconnections can persist with different probabilities (instead of with the common probability $\omega$). This would allow more flexibility for accurately capturing both the average contact and intercontact durations in real systems. We note that a mathematical analysis of such a generalization does not appear straightforward. Further, it is desirable to develop more sophisticated procedures for estimating the link persistence probabilities in real systems, e.g., based on maximum likelihood estimation. Also, it could be interesting to investigate the accuracy of the large-$N$ approximations [cf. Eqs.~(\ref{eq:approx1}) and (\ref{eq:approx2})] as a function of network sparsity ($\bar{k}/N$). Furthermore, it would be nice to investigate generalizations of the model that would allow the nodes' latent variables $(\kappa, \theta)$ to change over time (in the simplest case, via jump or walk dynamics as in Ref.~\cite{hartle2021}) and analyze the effect of the latent variables' motion on the resulting (inter)contact distributions and other temporal network properties. Finally, it would be interesting to investigate the exact effects of link persistence on spreading processes and related measures, such as the ones considered in Refs.~\cite{vazquez2007, Karsai2011}.

Taken altogether, our results advance our understanding of the realistic modeling of temporal networks with RHGs and of the effects of link persistence on temporal network properties. In addition to their explanatory power, parsimonious models, like the $\omega$-dynamic-$\mathbb{S}^1$, are also important for applications as they can constitute the basis of maximum likelihood estimation methods that more realistically infer the node coordinates and their evolution in the latent spaces of real systems~\cite{KimSurvey}. 

\section*{Acknowledgements}
We thank M. A. Rodr\'{i}guez-Flores for preparing the Email-EU data. S. Z. and F. P. acknowledge support by the TV-HGGs project (OPPORTUNITY/0916/ERC-CoG/0003), co-funded by the European Regional Development Fund and the Republic of Cyprus through the Research and Innovation Foundation. H. H. acknowledges support from NSF grant IIS-1741355.

\appendix

\section{Evaluating the integral in Equation~(\ref{eq:lim1_uncon})} 
\label{sec:Appendix1}

Evaluating the integral in~(\ref{eq:lim1_uncon}) using Mathematica~\cite{Mathematica} yields
\begin{align}
\label{eq:lim1} 
\nonumber & I_1 \coloneqq \int \limits_0^1 u^{-T}  (1 - u)^{1+T} [1-\xi (1-u)] ^{t-1} \mathrm{d}u \\
\nonumber &=\frac{T \pi (1-\xi)^{t-1}}{\xi \sin{(T \pi) (t+1)}}\Big\{(t-1) {}_2 F_{1} [2-t, 1-T; 2; \frac{\xi}{\xi-1}]\\
& -[t-1-\xi(t+T)] {}_2 F_{1} [1-t, 1-T; 2; \frac{\xi}{\xi-1}] \Big\}.
\end{align}
Below, we show that~(\ref{eq:lim1}) can be simplified, leading to~(\ref{eq:lim1_simpler}).

We first recall that the hypergeometric function is defined by the Gauss series
\begin{equation}
\label{eq:hypeq}
{}_2 F_{1}[a, b; c; z]=\sum_{n=0}^\infty \frac{(a)_n (b)_n}{(c)_n} \frac{z^n}{n!},
\end{equation}
for $|z| < 1$, and by analytic continuation elsewhere~\cite{special_functions_book}. The symbol $(q)_n$ is the  Pochhammer symbol, defined as $(q)_n=1$ for $n=0$, and $(q)_n=q (q+1) \ldots (q+n-1)$ for $n > 0$. Further, the following identity holds, known as Pfaff's transformation (Eq.~15.8.1 in Ref.~\cite{special_functions_book}):
\begin{equation}
\label{eq:Pfaff1}
{}_2 F_{1}[a, b; c; z] = (1-z)^{-a} {}_2 F_{1}[a, c-b; c; \frac{z}{z-1}].
\end{equation}

Using~(\ref{eq:Pfaff1}) for $(a, b, c, z) = (2-t, 1+T, 2, \xi)$ gives
\begin{equation}
\label{eq:P1} 
{}_2 F_{1} [2-t,1-T; 2; \frac{\xi}{\xi-1}] =  (1-\xi)^{2-t} {}_2 F_{1} [2-t,1+T; 2; \xi].
\end{equation}
Also, using~(\ref{eq:Pfaff1}) for $(a, b, c, z)=(1-t, 1+T, 2, \xi)$ gives
\begin{equation}
\label{eq:P2}
 {}_2 F_{1}[1-t, 1-T; 2; \frac{\xi}{\xi-1}] = (1-\xi)^{1-t}  {}_2 F_{1}[1-t, 1+T; 2; \xi].
\end{equation}
Now, using~(\ref{eq:P1}) and~(\ref{eq:P2}), we can rewrite~(\ref{eq:lim1}) as 
\begin{align}
\label{eq:lim1b} 
\nonumber I_1=\frac{T\pi}{\xi \sin{(T \pi)} (t+1)} \Big\{(t-1) (1-\xi) {}_2 F_{1} [2-t,1+T; 2; \xi]\\
-[t-1-\xi(t+T)] {}_2 F_{1}[1-t, 1+T; 2; \xi]\Big\}.
\end{align}

Equation~(\ref{eq:lim1b}) can be simplified by utilizing two of Gauss's relations between contiguous hypergeometric functions, namely, Eqs.~(34) and (42) in Sec.~2.8 of Ref.~\cite{bateman1953higher}, shown below,  
\begin{align}
\label{eq:F1} 
\nonumber  c [a-(c-b)z] {}_2 F_{1}[a, b; c; z] - a c (1-z) {}_2 F_{1} [a+1,b; c; z]\\
+ (c-a)(c-b)z {}_2 F_{1} [a, b; c+1; z] = 0,
\end{align}
and
\begin{align}
\label{eq:F2} 
\nonumber (c-b-1) {}_2 F_{1} [a, b; c; z] + b {}_2 F_{1}[a, b+1; c; z]\\
 -(c-1) {}_2 F_{1}[a, b; c-1; z] = 0.
\end{align}
Specifically, using~(\ref{eq:F1}) with $a=1-t, b=1+T, c=2$, and $z=\xi$, we can write
\begin{align}
\label{eq:F3}
\nonumber [1-t-\xi(1-T)] {}_2 F_{1} [1-t, 1+T; 2; \xi] + (t-1) \\
\nonumber \times (1-\xi) {}_2 F_{1} [2-t, 1+T; 2; \xi] \\
 = - \frac{1}{2} (t+1)(1-T)\xi {}_2 F_{1} [1-t, 1+T; 3; \xi].
\end{align}
Also, using~(\ref{eq:F2}) with $a=1-t, b=1+T, c=3$, and $z=\xi$, we have
\begin{align}
\label{eq:F4}   
\nonumber \frac{1+T}{2} {}_2 F_{1} [1-t, 2+T; 3; \xi] = {}_2 F_{1} [1-t, 1+T; 2; \xi]\\
- \frac{1-T}{2} {}_2 F_{1} [1-t,1+T; 3; \xi].
\end{align}
Now, from~(\ref{eq:F3}) we can rewrite~(\ref{eq:lim1b}) as
\begin{align}
\label{eq:lim1c} 
\nonumber I_1&=\frac{T\pi}{\sin{(T \pi)}} \Big\{{}_2 F_{1}[1-t, 1+T; 2; \xi]\\
&-\frac{1-T}{2} {}_2 F_{1} [1-t, 1+T; 3; \xi]\Big\}.
\end{align}
Further, from~(\ref{eq:F4}) we can simplify~(\ref{eq:lim1c}) to
\begin{equation}
\label{eq:lim1d} 
I_1=\frac{T (1+T) \pi}{2 \sin{(T \pi)}} {}_2 F_{1} [1-t, 2+T; 3; \xi].
\end{equation}
Using the above relation in~(\ref{eq:lim1_uncon}), and noticing that ${}_2 F_{1} [a, b; c; z]={}_2 F_{1} [b, a; c; z]$, yields~(\ref{eq:lim1_simpler}).

\section{Evaluating the integral in Equation~(\ref{eq:lim2_uncon})} 
\label{sec:Appendix2}

Evaluating the integral in~(\ref{eq:lim2_uncon}) using Mathematica~\cite{Mathematica} yields
\begin{align}
\label{eq:lim2} 
\nonumber I_2 &\coloneqq \int \limits_0^1 u^{1-T}  (1 - u)^T (1-\xi u)^{t-1} \mathrm{d}u\\
\nonumber & =\frac{(1-T)\pi}{\xi \sin{(T \pi) (t+1)}} \Big\{(1-\xi) {}_2 F_{1} [1-t, 2-T; 1; \xi]\\
&+[\xi(t+T)-1] {}_2 F_{1} [1-t, 2-T; 2; \xi] \Big\}.
\end{align}
Equation~(\ref{eq:lim2}) can be simplified, leading to~(\ref{eq:lim2_simpler}). 

To this end, we again utilize two of Gauss's relations between contiguous hypergeometric functions, namely, Eqs.~(43) and~(44) in Sec.~2.8 of Ref.~\cite{bateman1953higher}, shown below,
\begin{align}
\label{eq:F5}
\nonumber c (1-z) {}_2 F_{1} [a, b; c; z] - c {}_2 F_{1} [a, b-1; c; z]  \\
+ (c-a) z  {}_2 F_{1} [a, b; c+1; z] = 0,
\end{align}
and
\begin{align} 
\label{eq:F6} 
\nonumber [b-1-(c-a-1)z] {}_2 F_{1} [a, b; c; z] + (c-b)\\
\times {}_2 F_{1} [a, b-1;c;z] - (c-1)(1-z) {}_2 F_{1} [a,b;c-1;z] = 0.
\end{align}
Using~(\ref{eq:F6}) with $a=1-t, b=2-T, c=2$, and $z=\xi$ gives
\begin{align}
\label{eq:F7} 
\nonumber (1-\xi) {}_2 F_{1} [1-t, 2-T; 1; \xi]  = (1-T-\xi t )\\
\times {}_2 F_{1} [1-t, 2-T; 2; \xi] + T {}_2 F_{1} [1-t,1-T; 2; \xi].
\end{align}
Also, using~(\ref{eq:F5}) with $a=1-t, b=2-T, c=2$, and $z=\xi$ gives
\begin{align}
\label{eq:F8} 
\nonumber (1-\xi) {}_2 F_{1} [1-t, 2-T; 2; \xi]-{}_2 F_{1} [1-t, 1-T; 2; \xi] =  \\
= -\frac{(t+1)\xi}{2} {}_2 F_{1} [1-t, 2-T; 3; \xi].
\end{align} 

Now, from~(\ref{eq:F7}) we can rewrite (\ref{eq:lim2}) as
\begin{align}
\label{eq:lim2a} 
\nonumber I_2 & =\frac{T (1-T)\pi}{\xi \sin{(T \pi)} (t+1)} \Big\{(\xi-1) {}_2 F_{1} [1-t, 2-T; 2; \xi]\\
&+ {}_2 F_{1} [1-t,1-T;2;\xi] \Big\}.
\end{align}
Further, from~(\ref{eq:F8}) we can simplify~(\ref{eq:lim2a}) to
\begin{equation}
\label{eq:lim2b} 
I_2  =\frac{T (1-T)\pi}{2 \sin{(T \pi)}} {}_2 F_{1} [1-t, 2-T; 3; \xi].
\end{equation}
Using the above relation in~(\ref{eq:lim2_uncon}), and the fact that ${}_2 F_{1} [a, b; c; z]={}_2 F_{1} [b, a; c; z]$, yields~(\ref{eq:lim2_simpler}).


%

\end{document}